\documentclass[aps,pra,reprint,longbibliography,nofootinbib]{revtex4-2}

\usepackage{amsmath,amssymb,mathtools,bm}
\usepackage{siunitx}
\usepackage{microtype}
\usepackage{graphicx}
\usepackage{booktabs}
\usepackage[dvipsnames]{xcolor}
\usepackage{hyperref}
\usepackage[nameinlink,capitalise]{cleveref}

\hypersetup{colorlinks=true,linkcolor=MidnightBlue,citecolor=MidnightBlue,urlcolor=MidnightBlue}

\newcommand{\ket}[1]{\left|#1\right\rangle}
\newcommand{\bra}[1]{\left\langle#1\right|}
\newcommand{\braket}[2]{\left\langle#1\middle|#2\right\rangle}
\newcommand{\avg}[1]{\left\langle #1 \right\rangle}

\newcommand{\Tr}{\mathrm{Tr}}

\graphicspath{{figs/}{./}}

\begin{document}

\title{Angular Gausslets}

\author{Steven R. White}
\email{srwhite@uci.edu, corresponding author}
\affiliation{Department of Physics and Astronomy, University of California, Irvine, Irvine, CA 92697, USA}
\date{\today}

\begin{abstract}
Gausslets are one of the few basis constructions for electronic structure that combine locality,
orthonormality, variable resolution, and an accurate diagonal approximation for the
electron-electron interaction, but the original construction is tied to one dimension. Radial
gausslets extended this idea to atoms while leaving the angular degrees of freedom in spherical
harmonics, so the atomic interaction remained only partially diagonal in the combined basis.
Here we introduce generalized gausslets on the sphere and combine them shell by shell with
radial gausslets to form an atom-centered basis in which the electron-electron interaction takes
a two-index integral-diagonal form. The angular basis starts from localized spherical Gaussians
and uses injection to make a low-$\ell$ spherical-harmonic subspace exact. Tests of the kinetic
spectrum, low-$\ell$ Coulomb matrix elements, spherium, first-row Hartree--Fock calculations,
and He exact diagonalization show systematic convergence with increasing angular resolution. We
also develop DMRG methods for this basis, including compact MPOs, correlated small-space
starting states, Givens-rotation transfers between nearby angular sizes, and embedded sampled
variance extrapolation (ESVE). 
We show that this combination of ingredients can be used to solve the Be atom, with extrapolations in the number of angular functions but with fixed radial resolution,
to within about 0.1 mH of the complete basis set limit exact energy. This shows that DMRG
calculations of first row atoms which include both static and accurate
dynamic correlation on the same footing are feasible. 
\end{abstract}

\maketitle

\section{Introduction and background}
\label{sec:intro}

Gausslets\cite{White17,WhiteStoudenmire2019MultiSliced,QiuWhite2021Hybrid,White2023} are local orthogonal basis functions for electronic structure with an unusual combination of properties: orthonormality, locality, smoothness, variable resolution, and an accurate diagonal approximation for the two-electron interaction. While sitting alongside other localized or systematically improvable real-space constructions, including discrete-variable and related grid/quadrature methods,\cite{Light85,ColbertMiller1992DVR} multiresolution approaches,\cite{mrchem} and B-spline bases for atomic and continuum problems,\cite{Bachau2001BSplines,Zatsarinny06}, this combination of gausslet properties is so far unique.
Their main limitation is geometric. The original construction is only known in one dimension, so three-dimensional applications have relied on coordinate-product bases built from 1D ingredients.

For atoms, a Cartesian product structure is not the natural starting point. In Ref.~\cite{White2026RadialGausslets} we introduced radial gausslets, replacing the radial coordinate by a compact localized basis while keeping the angular dependence in spherical harmonics. That gave a useful atomic basis and a partial diagonal approximation: the interactions in the radial direction were diagonal, but the angular components were not. 

Here we develop the missing angular piece. We construct generalized gausslets on the sphere from localized spherical Gaussians and then inject an exact low-$\ell$ spherical-harmonic subspace.  When this angular basis is combined with radial gausslets, one has a fully diagonal interaction. We retain the full Galerkin form for the one-electron part of the Hamiltonian, for high accuracy and modest cost. This is the same Hamiltonian form as for Cartesian gausslets, but with a more natural representation for atoms. Importantly, the angular construction is much more general than the 1D wavelet construction, potentially allowing generalization to direct construction in 3D.

Fully diagonal interactions are particularly useful for certain correlated calculations. In standard Gaussian-basis QC-DMRG, the generic four-index interaction is a major bottleneck.\cite{WoutersVanNeck2014,SteinReiher2016,BaiardiReiher2020,Brabec2021,ZhaiChan2021} Consequently, DMRG is typically used only for active spaces or small bases. Here that part of the Hamiltonian is strongly compressed, allowing a much larger basis that includes dynamic correlation. The case of atoms is much harder for DMRG than the hydrogen-chain problems studied in earlier gausslet work\cite{StoudenmireWhite2017,WhiteStoudenmire2019,SawayaWhite2022,QiuWhite2021,WhiteLindsey2023}, which have one energy scale and a 1D arrangement. We therefore also develop practical DMRG methods for this basis: compact MPO representations, a correlated small-space starting calculation, Givens-rotation transfers between nearby angular sizes, and embedded sampled variance extrapolation (ESVE) for larger target spaces.

This paper is organized as follows: first, in Sec.~\ref{sec:ida} we explain the integral diagonal approximation that motivates the construction, how it can be formulated in a more general fashion than previously done, and the basis properties it requires. In Sec.~\ref{sec:basis} we then develop the generalized-gausslet construction in one dimension, introduce optimized point sets on the sphere, and build the  angular basis on the sphere. In Sec.~\ref{sec:hamiltonian-benchmarks} we couple that angular basis to radial gausslets, obtaining a 3D basis with a full diagonal approximation, and test the resulting Hamiltonian through angular diagnostics, first-row Hartree--Fock calculations, and exact diagonalization for He. Finally, in Sec.~\ref{sec:mpsbe} we apply DMRG and the associated technology to the Be atom in this basis, where exact diagonalization is no longer practical in the larger bases; Sec.~\ref{sec:discussion} summarizes and concludes. 

\section{The integral diagonal approximation}
\label{sec:ida}
Perhaps the most important property of gausslets is their support of diagonal
approximations for the two-electron interaction, turning a four-index tensor
into a two-index matrix. The most useful form defining the diagonal
approximation is the integral diagonal approximation (IDA). In previous work,
the IDA has been 
justified based on moment properties of the gausslets: the gausslets integrate
smooth functions like $\delta$-functions. In this section we show that the IDA
has a more general origin, at least to low order, stemming from completeness,
locality, and orthogonality, but not on moments. The accuracy of the IDA
without moment properties depends on the details of the basis; essentially,
having a basis that imitates gausslets in properties such as strong locality
and even symmetry tends to work well. We will first derive the IDA in this more
general fashion, and then discuss improving the accuracy. Since the key
application here is angular gausslets, we derive IDA on the sphere, but the
derivation naturally generalizes to other dimensions. For simplicity we start
with the one-particle IDA.

Consider an orthonormal localized angular basis $\phi_a(\Omega)$ on the sphere. For any one-particle multiplication operator $U(\Omega)$, the exact Galerkin matrix is
\begin{equation}
U_{ab} = \int d\Omega\, \phi_a(\Omega)\,U(\Omega)\,\phi_b(\Omega).
\label{eq:Uabexact}
\end{equation}
Expand
a smooth wavefunction as
\begin{equation}
\psi(\Omega) = \sum_b \psi_b \phi_b(\Omega).
\label{eq:psiExpand}
\end{equation}
Because of orthonormality, 
\begin{equation}
\psi_b = \int d\Omega\, \phi_b(\Omega) \psi(\Omega) .
\label{eq:psicoeff}
\end{equation}
Because the $\phi_a$ are localized, the exact matrix $U_{ab}$ is effectively
short-ranged in the basis index: for fixed $a$, only basis functions $b$ centered
near $a$ couple significantly. Define the weight of a basis function as
\begin{equation}
w_a = \int d\Omega\,\phi_a(\Omega).
\label{eq:angweight}
\end{equation}
Gausslets resemble positive bumps with short oscillatory tails, so we will assume $w_a > 0$. 
If $\psi(\Omega)$ varies slowly over that local
neighborhood, then throughout the support of the relevant basis functions it is
approximately constant, say $\psi(\Omega)\approx c_a$.  It follows that
\begin{equation}
\psi_b \approx c_a\, w_b,
\qquad
\psi_a \approx c_a\, w_a,
\qquad
c_a \approx \frac{\psi_a}{w_a}.
\label{eq:psismooth}
\end{equation}
Hence the action of $U$ on $\psi$ is
\begin{equation}
(U\psi)_a = \sum_b U_{ab}\psi_b
\approx
\left(\frac{1}{w_a}\sum_b U_{ab} w_b\right)\psi_a.
\label{eq:UDA1}
\end{equation}
This defines an effective diagonal one-particle operator
\begin{equation}
\widetilde U_{aa} \equiv \frac{1}{w_a}\sum_b U_{ab}w_b.
\label{eq:UDA2}
\end{equation}

We now assume that the basis represents the constant function exactly, so that
\begin{equation}
1 = \sum_b w_b\,\phi_b(\Omega).
\label{eq:constantfit}
\end{equation}
Substituting this into Eq.~(\ref{eq:UDA2}) gives
\begin{align}
\sum_b U_{ab}w_b
&=
\sum_b \int d\Omega\, \phi_a(\Omega)\,U(\Omega)\,\phi_b(\Omega)\,w_b
\nonumber\\
&=
\int d\Omega\, \phi_a(\Omega)\,U(\Omega)\,
\left(\sum_b w_b\phi_b(\Omega)\right)
\nonumber\\
&=
\int d\Omega\, \phi_a(\Omega)\,U(\Omega).
\label{eq:UDA3}
\end{align}
Therefore
\begin{equation}
\boxed{
\widetilde U_{aa}
=
\frac{1}{w_a}\int d\Omega\, \phi_a(\Omega)\,U(\Omega).
}
\label{eq:IDAonebody}
\end{equation}
This is the integral diagonal approximation (IDA) for a one-particle local operator.
Its derivation uses only locality, smoothness of $\psi$ over the coupling range, and
the exact representation of the constant function.  

Ordinary gausslets obey high order moment conditions, which result in the IDA
working to high order. The present derivation is explicitly low order, and so
the usefulness of this IDA without moments must be carefully validated, once we
have candidate basis sets. 

The IDA is used in practice only for the two-electron interaction, and its generalization to that case follows essentially from applying the one particle derivation to both angular coordinates.
For basis functions $\phi_a(\Omega)$ on a sphere of radius $R$, the exact four-index angular Coulomb integrals are
\begin{equation}
(ab|cd)
=
\int d\Omega\,d\Omega'\,
\phi_a(\Omega)\phi_b(\Omega)\,
\frac{1}{|R\Omega-R\Omega'|}\,
\phi_c(\Omega')\phi_d(\Omega'),
\label{eq:eriSphere}
\end{equation}
where inside the absolute value the directions $\Omega$ are regarded as unit vectors. 
In the integral diagonal approximation, the pair densities on each electron coordinate
are collapsed to diagonal form,
\begin{equation}
(ab|cd) \approx \delta_{ab}\,\delta_{cd}\,V_{ac},
\label{eq:twobodyIDA}
\end{equation}
with the two-index interaction matrix
\begin{equation}
V_{ac}
=
\frac{1}{w_a w_c}
\int d\Omega\,d\Omega'\,
\phi_a(\Omega)\,
\frac{1}{|R\Omega-R\Omega'|}\,
\phi_c(\Omega').
\label{eq:Vacdef}
\end{equation}
Thus the interaction is represented by $O(N^2)$ numbers rather than a full four-index
tensor.

For later use, we consider briefly here how one calculates $V_{ac}$ in practice, in terms of spherical harmonics. The Coulomb kernel has the multipole
expansion
\begin{equation}
\frac{1}{|R\Omega-R\Omega'|}
=
\frac{4\pi}{R}
\sum_{L=0}^\infty \frac{1}{2L+1}
\sum_{M=-L}^{L}
Y_{LM}(\Omega)\,Y_{LM}^*(\Omega').
\label{eq:CoulSphere}
\end{equation}
Using the moments
\begin{equation}
M_{LM}^{(a)} \equiv \int d\Omega\, Y_{LM}(\Omega)\,\phi_a(\Omega),
\label{eq:MLMop}
\end{equation}
Eq.~(\ref{eq:Vacdef}) becomes
\begin{equation}
V_{ac}
=
\frac{4\pi}{R\,w_a w_c}
\sum_{L=0}^\infty \frac{1}{2L+1}
\sum_{M=-L}^{L}
M_{LM}^{(a)}\,M_{LM}^{(c)},
\label{eq:Vacmoments}
\end{equation}
where for real spherical harmonics the complex conjugation is omitted.
In practice the sum over $L$ is truncated when the last shell is numerically negligible.
Equation~(\ref{eq:Vacmoments}) allows us to assemble
the Coulomb interaction from a two-index matrix built from
the moments of the localized basis functions, without ever forming the full four-index
tensor.

\section{Basis construction}
\label{sec:basis}

The goal of the basis construction is to build a localized basis that satisfies the conditions identified in Sec.~\ref{sec:ida}, especially exact constant-function completeness,  orthonormality, and sharp locality. In addition, we also desire additional low momentum completeness, beyond representing a constant. As a guiding principle, we try to imitate both the properties and construction of gausslets, but in a more general fashion. We first discuss a simple one-dimensional
version, in order to show most clearly how it works. Later, we introduce the spherical geometry and atomic Hamiltonians. 

\subsection{Generalized gausslets in 1D}
\label{sec:generalized-gausslets}

In the original gausslet construction\cite{White17}, a uniform array of Gaussians is
transformed by a wavelet step and a convolution into orthonormal localized gausslets. That
construction is specific to one dimension. Here we use a simpler starting point that keeps the
same basic idea and can be generalized more easily. Consider equally spaced identical
Gaussians. If they are made wide at fixed spacing, they represent low-momentum modes very
well,\cite{White17} but the overlap matrix becomes nearly singular and symmetric
orthogonalization produces long tails, approaching sinc functions in the extreme limit. If
they are made narrow, the overlap matrix is well conditioned and the orthogonalized functions
are very local, but the low-momentum completeness is poor. Between these limits there is a
useful intermediate regime: the overlap matrix is only moderately ill-conditioned, the
orthogonalized functions remain local, and the low-momentum completeness is already good.

\begin{figure}
  \centering
  \includegraphics[width=0.6\columnwidth]{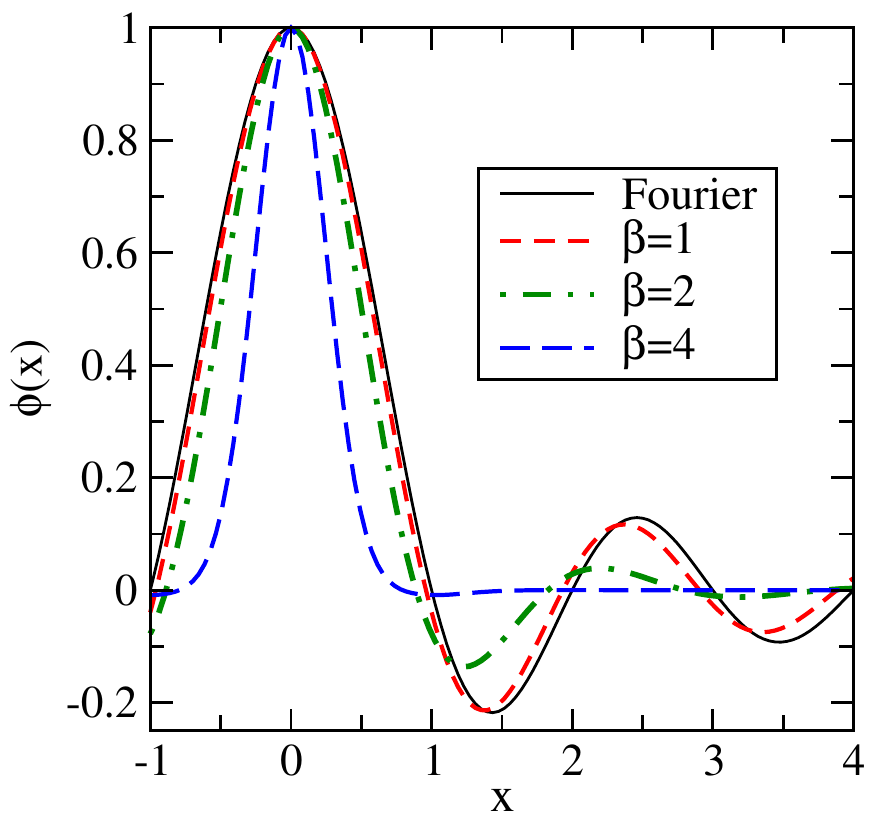}
  \caption{Centered orthonormalized Gaussian functions on the periodic 1D line for
  $\beta=1,2,4$, compared with the Fourier-localized packet formed from the same number of
  plane waves.  The Fourier-localized function has long range oscillatory tails, while the
  orthonormalized Gaussians are much more local. In this test calculation, the length of the periodic line was 41, long enough to clearly show tails.}
  \label{fig:1dlocality}
\end{figure}

To illustrate this, consider a finite but long periodic 1D line with unit-spaced centers and one
Gaussian $\exp(-\beta^2 x^2/2)$ of width $1/\beta$ on each site. After symmetric (L\"owdin) orthogonalization\cite{Lowdin1950}
the basis is translationally invariant, so it is enough to examine the
function centered at the origin. Figure~\ref{fig:1dlocality} compares that function for several
widths with the standard Fourier-localized packet built from the same number of plane waves.
The Fourier-localized packet has long oscillatory tails. The $\beta=1$ orthonormalized
Gaussian still has noticeable tails, the $\beta=4$ function is very sharp, and $\beta=2$
lies in between with good localization and only short-range tails.

\begin{figure}
  \centering
  \includegraphics[width=0.6\columnwidth]{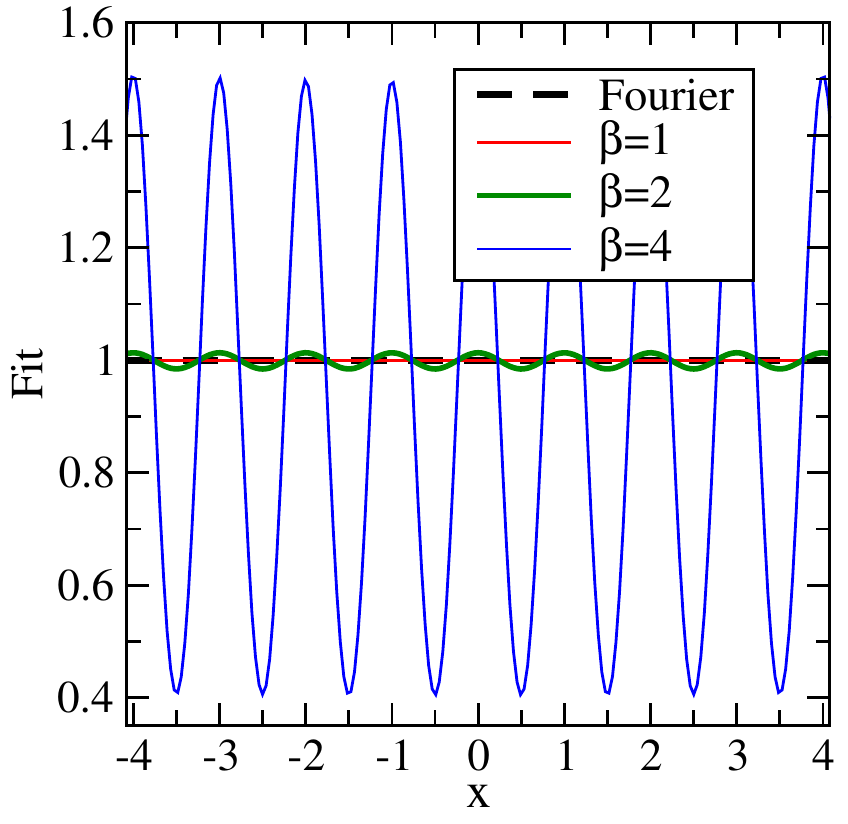}
  \caption{Best fit to the constant function within the orthonormalized Gaussian span.
  The Fourier curve is exactly constant.  The narrow $\beta=4$ Gaussians show strong
  oscillatory errors in the constant fit, while $\beta=2$ and especially $\beta=1$
  represent the low-momentum sector much more accurately.}
  \label{fig:1dconstant}
\end{figure}

In Figure~\ref{fig:1dconstant} we see how well the functions fit a constant, which is a good indicator for Gaussians of low-momentum completeness, and essential for the IDA. The Fourier fit is perfect, for $\beta=1$ it is excellent, for $\beta=2$ it is good, and for $\beta=4$ it is terrible.  The condition number of the overlap matrix is $8700$, $5.9$, and $1.1$ for $\beta=1,2,4$.  Thus there is an intermediate regime---the sweet
spot---where the functions are still nicely localized, the overlap matrix is not too singular, and the low-momentum completeness is
already quite good.  For this example, $\beta \approx 2$, or a little less, is in that regime.

Good low-momentum completeness is still not enough. Even small errors in the lowest momentum
modes show up directly in energies and in diagonal approximations. We therefore modify the
basis by injection: we replace the approximate low-momentum part of the Gaussian span by an
exact low-momentum subspace, and then orthonormalize within that new span. More precisely, let
$G_i(x)$ denote the orthonormalized Gaussian basis, with $G$ its span, and let $Y$ denote the
low-momentum subspace we want to represent exactly (for example, the plane waves with
$|k|\le k_{\rm inj}$), with $Y^\perp$ its complement.
We define the \emph{injected span} by
\begin{equation}
F \equiv Y \oplus (G \cap Y^\perp).
\label{eq:injectedspan}
\end{equation}
That is, the approximate low-momentum content of $G$ is replaced by the exact
subspace $Y$, while all remaining directions are taken from the part of $G$
orthogonal to $Y$.

An equivalent operational definition can be given directly in terms of the basis
functions $G_i(x)$, $i=1\ldots N$.
Let $y_\mu(x)$, $\mu=1,\ldots,N_Y$ be the low-momentum functions to be injected, with $N_Y < N$.  The coefficients of the best fit of the $y_\mu$ in the $G$ basis are
\begin{equation}
C_{i\mu} = \int_0^L dx\, G_i(x)\,y_\mu(x)
\label{eq:yGproj}
\end{equation}
which is $N\times N_Y$. Form the orthogonal complement $B$ of this matrix, which is any $N\times (N-N_Y)$ matrix whose orthonormal columns are all orthogonal to the columns of $C$. Then the set of
\begin{equation}
q_\alpha(x) = \sum_i B_{i\alpha} G_i(x)
\label{eq:qalpha}
\end{equation}
spans $G \cap Y^\perp$.  
The $q_\alpha$ are orthogonal to the $y_\mu$ and the final basis is spanned by
\begin{equation}
F = Y \oplus (G\cap Y^\perp)
  = \mathrm{span}\{y_\mu(x), q_\alpha(x)\}.
\label{eq:injspanx}
\end{equation}

If one only cared about the span, the set $\{y_\mu,q_\alpha\}$ would suffice.
However, the $y_\mu(x)$ are delocalized plane waves, whereas we want a localized final basis.
To obtain this, we project each original Gaussian basis function $G_i(x)$ into the injected
span $F$.  Since the coefficients of
$G_i$ along the exact modes are just $C_{i\mu}$, and the coefficients along the complement
are $B_{i\alpha}$, we define
\begin{equation}
\overline G_i(x)
=
\sum_{\mu=1}^{N_Y} C_{i\mu}\,y_\mu(x)
+
\sum_{\alpha=1}^{N-N_Y} B_{i\alpha}\,q_\alpha(x).
\label{eq:Gbar}
\end{equation}
Equivalently, using Eq.~(\ref{eq:qalpha}),
\begin{equation}
\overline G_i(x)
=
\sum_{\mu=1}^{N_Y} C_{i\mu}\,y_\mu(x)
+
\sum_j (B B^T)_{ij}\,G_j(x).
\label{eq:Gbar2}
\end{equation}
Thus the $\overline G_i(x)$ are obtained directly as contractions of the original
localized basis functions $G_j(x)$ and the exact injected functions $y_\mu(x)$.

Because the sets $\{y_\mu\}$ and $\{q_\alpha\}$ are orthonormal and mutually orthogonal,
the overlap matrix of the projected basis is simply
\begin{equation}
\overline S_{ij}=
(C C^T + B B^T)_{ij}.
\label{eq:Sbar}
\end{equation}
The final injected basis is then obtained by a symmetric (L\"owdin) orthogonalization,
\begin{equation}
G_i^{\rm inj}(x)
=
\sum_{j=1}^N \overline G_j(x)\, [\overline S^{-1/2}]_{ji}.
\label{eq:Ginj}
\end{equation}
Equivalently, the final basis has the explicit contracted form
\begin{equation}
G_i^{\rm inj}(x)
=
\sum_{\mu} A_{\mu i}\,y_\mu(x)
+
\sum_j D_{ji}\,G_j(x),
\label{eq:Ginjcontracted}
\end{equation}
with
\begin{equation}
A = C^T \overline S^{-1/2},
\qquad
D = B B^T \overline S^{-1/2}.
\label{eq:ADdefs}
\end{equation}

\begin{figure}
  \centering
  \includegraphics[width=0.6\columnwidth]{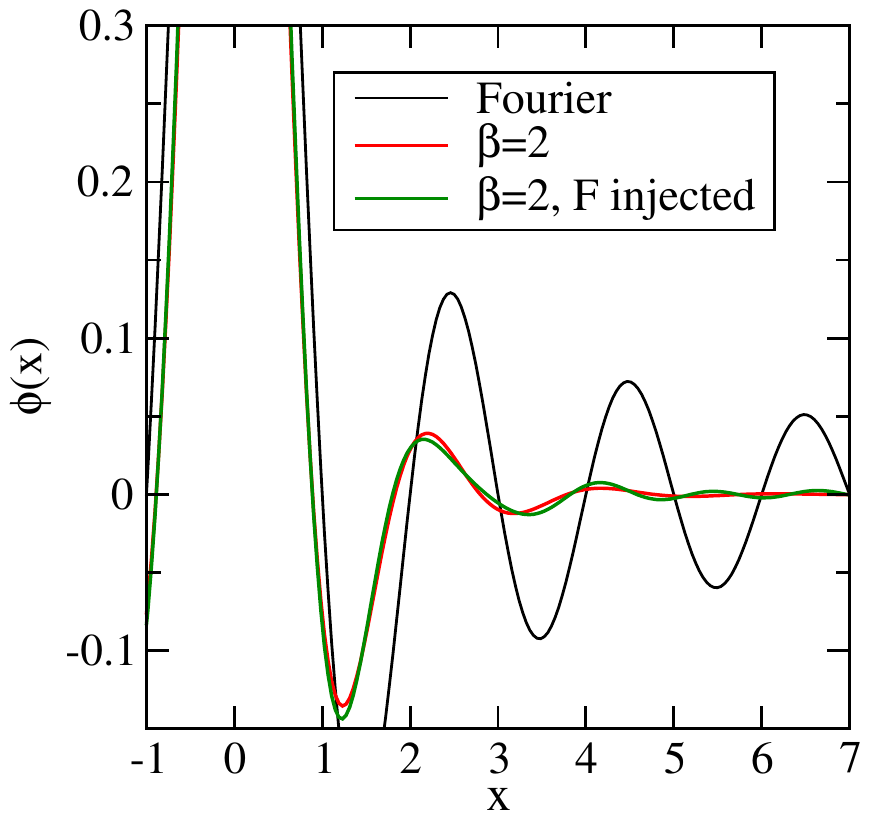}
  \caption{Effect of injection for the sweet-spot case $\beta=2$.  The injected localized
  function lies almost on top of the original orthonormalized Gaussian, while
  now containing the exact low-momentum content of the injected subspace.}
  \label{fig:1dinjection}
\end{figure}

Because the sweet-spot Gaussian basis already approximates the low-momentum sector well,
injection only changes the basis slightly. It makes the low-$k$ content exact while leaving the
localized functions nearly unchanged. Figure~\ref{fig:1dinjection} shows the result for the
sweet-spot case $\beta=2$. The injected function differs mainly in the tails. This is the
feature we need later on the sphere: a small correction to the smooth sector without
sacrificing locality.

The original gausslet construction utilizes $\beta=1$, more singular and more complete than the sweet spot. The
wavelet transformation together with an additional convolution bypass the singularity of the overlap matrix and generate orthonormal, very well
localized functions with the desired completeness and moment properties. Here the generalized gausslet procedure is less ideal, but it can be applied directly in higher dimensions.

In the following sections we apply this same logic on the sphere.  Plane waves are replaced
by spherical harmonics, the low-$k$ sector becomes the low-$\ell$ sector, and the injected
basis becomes a localized angular basis with exact low-$\ell$ content.

\subsection{Point sets on the sphere}
\label{sec:spherepoints}

On the line the centers are equally spaced. On the sphere the first question is how to choose
the centers $\hat n_i$ on $S^2$. Natural candidates are nearly uniform point sets such as
Lebedev grids,\cite{LebedevLaikov1999} spherical Fibonacci sets,\cite{Gonzalez2010Fibonacci} or spherical codes and other optimized distributions on the sphere.\cite{SaffKuijlaars1997} We first explored Lebedev points, since
they are familiar in electronic structure, but here they are not ideal: some available orders
are noticeably less uniform than one would like, and exact quadrature is not the main issue.
The exact low-$\ell$ sector will be enforced later by injection, but point selection affects completeness beyond that, and the conditioning of an associated overlap matrix. In general, we seek the most uniform point sets.

We associate to each point $\hat n_i$ a localized
spherical Gaussian centered in that direction, specifically
\begin{equation}
g_i(\Omega) = \exp\!\left[\kappa\,(\hat n_i\cdot \Omega - 1)\right],
\label{eq:sphgauss}
\end{equation}
where $\Omega$ specifies angular position via a unit vector. For the moment, all Gaussians have the same width parameter
$\kappa$. Their overlap matrix can be evaluated analytically:
\begin{equation}
S_{ij} = \int d\Omega\, g_i(\Omega) g_j(\Omega)
       = 4\pi e^{-2\kappa}\, \mathrm{sinhc}(R_{ij}),
\label{eq:sphoverlap}
\end{equation}
where
\begin{equation}
R_{ij} = \kappa \sqrt{2(1+\hat n_i\cdot \hat n_j)},
\qquad
\mathrm{sinhc}(x)=\frac{\sinh x}{x}.
\label{eq:sphRij}
\end{equation}
As in one dimension, the overlap matrix is the key object: if it is too close to
singular, orthogonalization produces long tails, while if the points are too
irregular the resulting localized basis is uneven and its completeness is compromised, and also its overlap matrix tends to become more singular.

We therefore introduce an optimization step, moving the points to optimize the conditioning of the overlap matrix directly, minimizing
\begin{equation}
f(\hat n_1,\ldots,\hat n_N) = -\log\det S.
\label{eq:logdetobj}
\end{equation}
This objective strongly penalizes small
eigenvalues of $S$, and hence near-linear dependence among the prototypes, while at
the same time favoring point sets which support a stable and reasonably uniform
localized basis after orthogonalization.  In practice this criterion works very
well: it gives center sets which are more useful for the present construction than
those chosen by quadrature considerations alone.

As starting points for the optimization, we use two simple and broadly available
families.  The first is the spherical Fibonacci construction, which gives a
deterministic point set for essentially any $N$.  The second, when available, is a
spherical code or packing set, which provides a very good initial guess for
uniformity\cite{SaffKuijlaars1997}.  For a given $N$, we optimize from both starts and keep whichever final
configuration gives the larger $\log\det S$.  This gives a practical and robust
route to a library of optimized point sets over a wide range of $N$.  The resulting
point sets are characterized only by their cardinality $N$; later, different radial
shells will use different values of $N$ according to the angular resolution needed
at that radius. Some cardinalities are also geometrically more favorable than
others; in particular, the $N=32$ case admits an unusually regular sphere
configuration with nearly uniform nearest-neighbor spacing, which helps explain
why it is especially clean in the diagnostics below. The GaussletBases.jl
software library noted at the end of this paper can provide the point sets.

The optimization just described uses a single global concentration parameter
$\kappa$ for all points, because at that stage we are only selecting the centers.
Once the centers are fixed, the actual localized basis uses widths determined from
the local nearest-neighbor spacing.  Denoting by $\theta_i^{\rm nn}$ the angular
distance from $\hat n_i$ to its nearest neighboring center, we take
\begin{equation}
\kappa_i = \frac{\beta^2}{(\theta_i^{\rm nn})^2+\epsilon},
\label{eq:kappann}
\end{equation}
with a small $\epsilon$ only to avoid numerical problems.  Thus the final width of
each prototype is tied to the local point density: where the centers are slightly
closer, the functions are slightly narrower, and vice versa.  This separates the
two tasks cleanly.  The point optimization chooses a good geometric scaffold for
localization and conditioning, while the local $\kappa_i$ adapt the basis smoothly
to the actual point set used in the production calculation.

These optimized point sets are not meant to
be used for angular quadratures. Their role in constructing angular gausslets is distinct. The exact
low-$\ell$ physics enters separately, through injection of the corresponding
$Y_{\ell m}$ subspace, as detailed in the next subsection. 

\subsection{Angular generalized gausslets on the sphere}
\label{sec:angbasis}

Given a set of spherical Gaussians
\begin{equation}
g_i(\Omega) = \exp\!\left[\kappa_i(\hat n_i\cdot \Omega - 1)\right],
\label{eq:sphproto}
\end{equation}
the next step is to inject an exact low-$\ell$ spherical-harmonic
subspace into it. Let
\begin{equation}
Y = \mathrm{span}\{Y_{\ell m}(\Omega): 0\le \ell \le L_{\rm inj}\},
\label{eq:Ysubspace}
\end{equation}
where the $Y_{\ell m}$ are taken orthonormal on the sphere. 
We choose $L_{\rm inj}$ as large as possible subject to 
$(L_{\rm inj}+1)^2 \le N/2$. 
The same injection construction described in Sec.~\ref{sec:generalized-gausslets} 
is then applied, with the 1D plane waves replaced by these spherical 
harmonics and with localization again obtained by 
projection onto the original Gaussians. 
After the final symmetric orthogonalization, we denote the resulting localized
orthonormal angular basis functions by $\phi_a(\Omega)$. These functions can be
written as contractions of the original Gaussians and the injected low-$\ell$
spherical harmonics.

For the final angular basis we use the same weights $w_a$ as in
Sec.~\ref{sec:ida}, and also define the first moment
of each angular orbital by
\begin{equation}
\mathbf m_a = \int d\Omega\,\Omega\,\phi_a(\Omega).
\label{eq:angweightdipole}
\end{equation}
We choose the phase of each $\phi_a$ so that $w_a\ge 0$. Additional locality and moment diagnostics, including the definitions of $c_{1,a}$, $\delta\gamma_a$, and $\epsilon_{\ell,a}$, are collected in Appendix~\ref{app:diagnostics}.

\begin{figure}
  \centering
  \includegraphics[width=0.8\columnwidth]{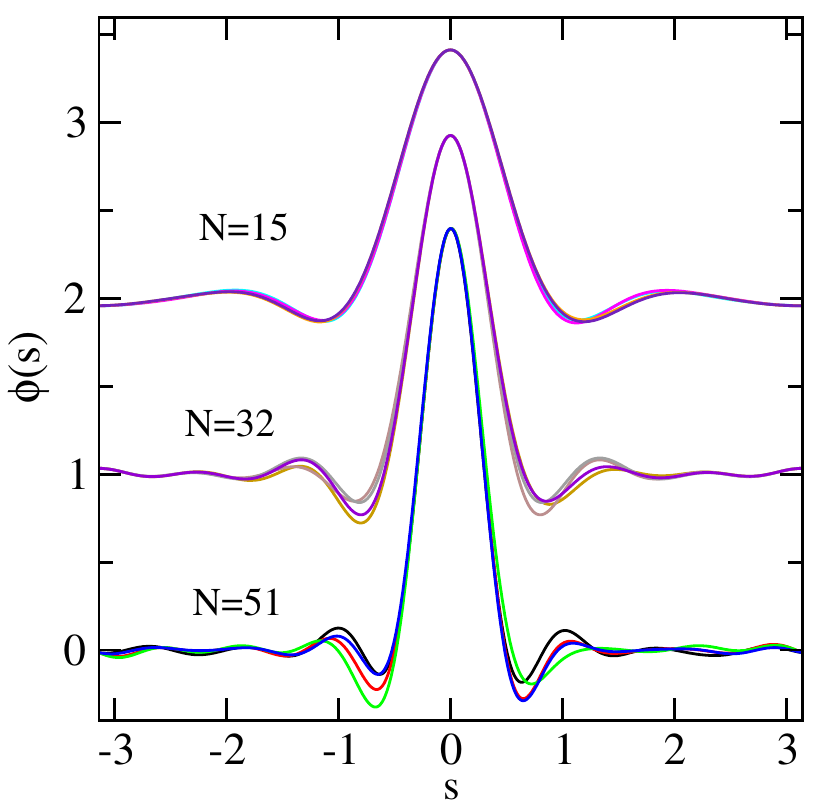}
  \caption{
Great-circle cuts through representative localized angular basis functions for optimized
sphere point sets with $N=15$, $32$, and $51$ centers (top to bottom). For the three point sets shown, the injected spherical-harmonic subspace was chosen as the largest complete low-$\ell$ block satisfying $(L_{\rm inj}+1)^2 \le N/2$, giving $L_{\rm inj}=1$ ($N_Y=4$) for $N=15$, $L_{\rm inj}=3$ ($N_Y=16$) for $N=32$, and $L_{\rm inj}=4$ ($N_Y=25$) for $N=51$.  For each $N$,
one final injected basis function is evaluated along four great circles through its
center, with tangent directions separated by $45^\circ$.  The horizontal axis $s$ is
the signed angular distance along the great circle, in radians, and the curves for
$N=32$ and $N=15$ are shifted upward by $1$ and $2$, respectively, for clarity.
}
  \label{fig:worldlines}
\end{figure}

Figure~\ref{fig:worldlines} shows one representative angular basis function for three values of
$N$. The four great-circle cuts almost lie on top of one another, so the functions depend
mainly on geodesic distance from their centers, with only weak directional anisotropy. They are
also strongly localized: each has one dominant peak, modest side lobes, and small tails. As
$N$ increases from 15 to 51 the peak sharpens and the oscillatory structure contracts, as
expected for a denser point set.

Appendix~\ref{app:diagnostics} gives quantitative locality and moment
diagnostics for several angular gausslet sets. The results are favorable, with
small weight variation and small higher-moment errors, but the IDA still has to
be tested directly in physical problems. In interacting calculations the
relevant question is whether the IDA error is below the basis-incompleteness
error at the same resolution.



\section{Hamiltonian construction and simple benchmarks}
\label{sec:hamiltonian-benchmarks}

We now test the Hamiltonians built from the basis of Sec.~\ref{sec:basis}. We
first examine purely angular problems on the sphere, and then
couple the angular basis to radial gausslets for atomic calculations.

\begin{figure}
  \centering
  \includegraphics[width=0.8\columnwidth]{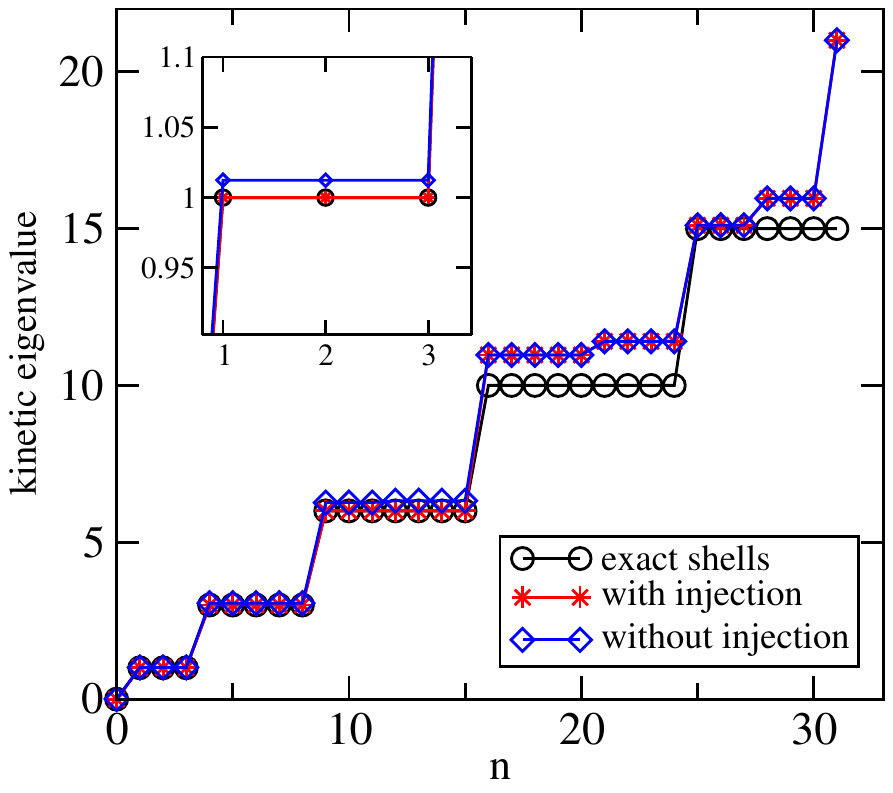}
  \caption{Kinetic-energy eigenvalues for the angular basis with $N=32$ optimized centers. This is a purely angular calculation on a sphere of radius $R=1$, with no radial degrees of
  freedom, corresponding to the one-particle angular Hamiltonian used in spherium.
  The black circles show the exact spherical-harmonic shell sequence
  $\ell(\ell+1)/2$ with degeneracies $2\ell+1$, ordered by eigenvalue index $n$.
  The red stars are the eigenvalues of the final injected basis, and the blue diamonds
  are those of the corresponding basis without injection.  For $N=32$ the injected
  subspace has $L_{\rm inj}=3$, so the first $(L_{\rm inj}+1)^2=16$ states, corresponding
  to the $\ell=0,1,2,3$ shells, should be exact.  This is seen directly: the injected
  basis reproduces those shells to numerical precision, while the non-injected basis
  already shows small errors in the lowest shells, visible most clearly in the inset for
  the $\ell=1$ triplet.  Deviations first appear beyond the injected block, beginning
  with the $\ell=4$ shell, as expected.}
  \label{fig:KE32}
\end{figure}

\subsection{Angular Hamiltonian tests and benchmarks}
\label{sec:tests}

A stringent test of the angular construction is whether the exact low-$\ell$ kinetic
spectrum is recovered.  Figure~\ref{fig:KE32} shows this for the $N=32$ basis, on a unit radius sphere with no radial degrees of freedom.
Because the injected subspace contains all spherical harmonics through $\ell=3$, the
first 16 kinetic eigenstates should agree exactly with the ideal shell sequence
$\ell(\ell+1)/2$.  They do: the injected basis is numerically exact through the
$\ell=3$ shell, while the basis without injection already shows small but visible
errors in the lowest shells.  Beyond the injected block, the eigenvalues deviate from
the exact shell values, which is expected since those higher-$\ell$ components are then
represented only by the localized Gaussian part of the basis.  Thus the figure shows
very directly that injection pins the low-angular-momentum physics exactly, while the
remaining basis provides an approximate but smooth continuation to higher $\ell$.

As a direct test of the two-index Coulomb construction, we compared the four-index
interaction tensor obtained from the integral diagonal approximation with the exact
four-index tensor in low-$\ell$ spherical-harmonic subspaces.  For each basis size $N$,
we restricted to the subspace spanned by all $Y_{\ell m}$ with $\ell \le L_0$, where
$L_0 \le L_{\rm inj}$ so that this test remains entirely within the exactly injected
low-$\ell$ sector.  We then formed the exact four-index Coulomb tensor in that subspace
and compared it with the tensor reconstructed from the two-index IDA interaction.  The
results are summarized in Table~\ref{tab:idaYlm} in Appendix~\ref{app:diagnostics}. The
overall conclusion is that the integral diagonal approximation is already accurate in the
low-$\ell$ sector most relevant to smooth angular physics, and that its quality improves
systematically with basis density. These diagnostics also support our default choice
$\beta=2$.

\begin{figure}
  \centering
  \includegraphics[width=0.8\columnwidth]{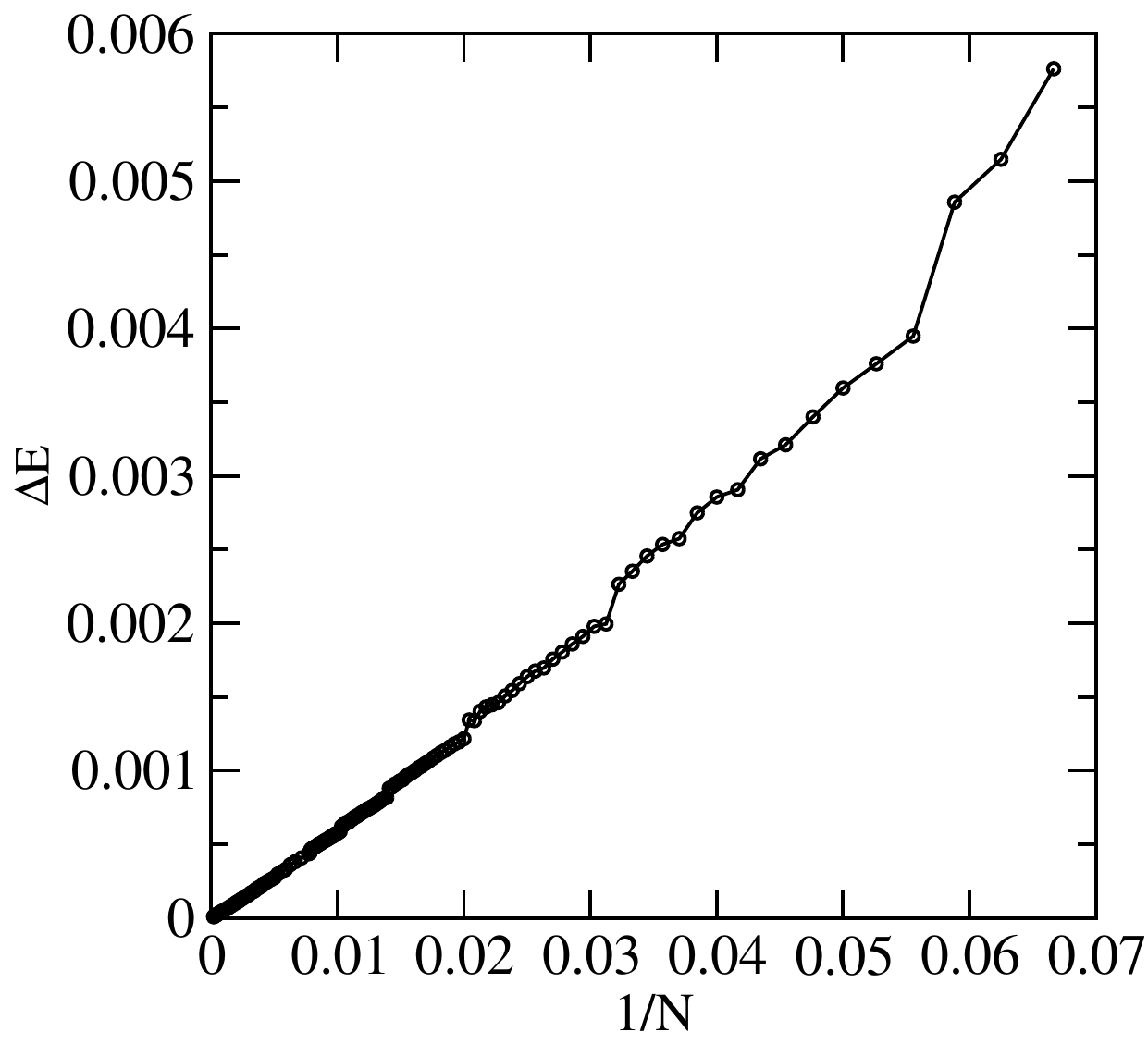}
\caption{Total energy errors in Hartrees for the spherium system of two opposite-spin electrons on a sphere of radius $R=1$, with ordinary kinetic energy and the standard Coulomb $1/r$ interaction between electrons.  The angular
kinetic operator is treated exactly, while the electron-
electron interaction is represented by the two-index IDA. The
plotted quantity is the ground-state energy error $\Delta E = E - E_{\mathrm{exact}}$ versus
$1/N$, where $N$ is the number of angular centers and $E_{\mathrm{exact}}=0.852781065$.  The systematic decrease of $\Delta E$ with increasing basis size, which is approximately linear in $1/N$ over the range shown, indicates fairly smooth convergence and suggests that the remaining error is dominated by basis
incompleteness rather than a breakdown of the two-index IDA.}
\label{fig:spherium}
\end{figure}

The standard spherium model\cite{LoosGill2010Spherium} provides a direct test of the purely angular Hamiltonian, complementary to the
one-body kinetic-spectrum test of Fig.~\ref{fig:KE32} and the low-$\ell$ Coulomb diagnostic
summarized in Appendix~\ref{app:diagnostics}. In this problem two electrons are constrained to move on a sphere
of radius $R=1$, so the calculation isolates the angular basis and the two-index Coulomb
construction without any radial degrees of freedom. Figure~\ref{fig:spherium} shows that the
ground-state energy converges systematically as the number of angular centers is increased.
Thus the present construction is not only exact in its injected low-$\ell$ sector and
accurate for low-$\ell$ Coulomb matrix elements, but is also effective in a full interacting
angular benchmark.

\subsection{Atomic basis and Hamiltonian construction}
\label{sec:atomicbasis}

We now combine the angular generalized gausslets with the radial gausslets to form an
atomic basis in three dimensions.  Let $\chi_a(r)$ denote the orthonormal radial
gausslets for the reduced radial coordinate, so that the physical one-electron basis
functions have the form
\begin{equation}
\varphi_{a\mu}(\mathbf r)
=
\frac{\chi_a(r)}{r}\,\phi_{\mu}^{(a)}(\Omega),
\qquad
\mathbf r \equiv (r,\Omega).
\label{eq:varphiamu}
\end{equation}
Here $\phi_\mu^{(a)}(\Omega)$ is the angular generalized-gausslet basis attached to
radial shell $a$.  Thus the angular basis is allowed to depend on the radial index:
different radial shells may use different point sets, different numbers of angular
functions, and different injected low-$\ell$ blocks.  This shell-dependent construction
is one of the main advantages of the present approach, since the required angular
resolution varies strongly with radius.  The basis remains orthonormal nevertheless:
the radial functions satisfy $\int_0^\infty \chi_a(r)\chi_b(r)\,dr=\delta_{ab}$, and
within each shell the angular functions are orthonormal,
$\int d\Omega\,\phi_\mu^{(a)}(\Omega)\phi_\nu^{(a)}(\Omega)=\delta_{\mu\nu}$.
Hence
\begin{equation}
\langle a\mu | b\nu \rangle = \delta_{ab}\delta_{\mu\nu}.
\label{eq:overlapfull}
\end{equation}
What changes, relative to a fixed $Y_{\ell m}$ basis, is that matrix elements between
different radial shells involve cross-shell angular overlaps and multipole couplings.

For the one-electron Hamiltonian
\begin{equation}
\hat h
=
-\frac{1}{2}\frac{d^2}{dr^2}
+
\frac{\hat L^2}{2r^2}
-
\frac{Z}{r},
\label{eq:h1atom}
\end{equation}
we use exact Galerkin matrix elements.  On the radial side we define
\begin{align}
T^r_{ab} &= \frac{1}{2}\int_0^\infty \chi_a'(r)\chi_b'(r)\,dr, \\
V^r_{ab} &= -Z\int_0^\infty \frac{\chi_a(r)\chi_b(r)}{r}\,dr, \\
C^r_{ab} &= \int_0^\infty \frac{\chi_a(r)\chi_b(r)}{r^2}\,dr.
\end{align}
On the angular side, because the basis may differ from shell to shell, we define the
cross-shell overlap and angular kinetic blocks
\begin{align}
S_{\mu\nu}^{\Omega,(a,b)}
&=
\int d\Omega\,\phi_\mu^{(a)}(\Omega)\phi_\nu^{(b)}(\Omega),
\label{eq:Sangcross}
\\
K_{\mu\nu}^{\Omega,(a,b)}
&=
\int d\Omega\,\phi_\mu^{(a)}(\Omega)\,
(-\Delta_\Omega)\,
\phi_\nu^{(b)}(\Omega).
\label{eq:Kangcross}
\end{align}
Then the exact one-electron matrix elements are
\begin{equation}
h_{a\mu,b\nu}
=
T^r_{ab}\,S_{\mu\nu}^{\Omega,(a,b)}
+
V^r_{ab}\,S_{\mu\nu}^{\Omega,(a,b)}
+
\frac{1}{2}\,C^r_{ab}\,K_{\mu\nu}^{\Omega,(a,b)}.
\label{eq:h1matrix}
\end{equation}
Thus the one-electron Hamiltonian is treated variationally and exactly in the chosen
basis, even though the angular functions vary from shell to shell.  In practice the
angular blocks are obtained directly from the injected $Y_{\ell m}$ part and the
Gaussian part of the localized angular functions, or equivalently from their spherical
moment tables.

The two-electron interaction is handled differently.  As in the radial-gausslet
construction, we do not form the full four-index Coulomb tensor.  Instead we use the
integral diagonal approximation and construct a two-index interaction matrix in the
product basis.  Let
\begin{equation}
w_a^r = \int_0^\infty \chi_a(r)\,dr
\label{eq:wrdef}
\end{equation}
denote the radial weights, and for each shell define the angular weights
\begin{equation}
w_\mu^{(a)} = \int d\Omega\,\phi_\mu^{(a)}(\Omega).
\label{eq:wangdef}
\end{equation}
For each multipole $L$, the radial kernel is
\begin{equation}
K_L(r,r')
=
\frac{r_<^L}{r_>^{L+1}},
\qquad
r_< = \min(r,r'),
\quad
r_> = \max(r,r'),
\label{eq:KLdef}
\end{equation}
and the corresponding radial IDA block is
\begin{equation}
V_{ab}^{(L),r}
=
\frac{1}{w_a^r w_b^r}
\int_0^\infty\!\!\int_0^\infty
\chi_a(r)\,K_L(r,r')\,\chi_b(r')\,dr\,dr'.
\label{eq:VradL}
\end{equation}
These are exactly the same radial multipole tables used in the radial-gausslet atomic
construction.

For the angular part, let
\begin{equation}
M_{LM,\mu}^{(a)}
=
\int d\Omega\,Y_{LM}(\Omega)\,\phi_\mu^{(a)}(\Omega)
\label{eq:MLMamu}
\end{equation}
be the multipole moments of the angular basis functions in shell $a$.  We then define
the inverse-weight-scaled moments
\begin{equation}
B_{M\mu}^{(L,a)}
=
\frac{1}{w_\mu^{(a)}}\,M_{LM,\mu}^{(a)},
\qquad
M=-L,\ldots,L,
\label{eq:BLadef}
\end{equation}
and from them the angular IDA blocks
\begin{equation}
A_{\mu\nu}^{(L),(a,b)}
=
\frac{4\pi}{2L+1}
\sum_{M=-L}^{L}
B_{M\mu}^{(L,a)}\,B_{M\nu}^{(L,b)}.
\label{eq:ALabdef}
\end{equation}
For $L=0$, injection of $Y_{00}$ makes the constant function exact, and one obtains
$B_{0\mu}^{(0,a)}=1/\sqrt{4\pi}$ for every shell and orbital.  Thus the monopole block
is reproduced exactly,
\begin{equation}
A_{\mu\nu}^{(0),(a,b)} = 1,
\label{eq:A0exact}
\end{equation}
which provides a useful numerical check on the implementation.

The final two-index Coulomb interaction in the atomic product basis is therefore
\begin{equation}
V_{a\mu,b\nu}
=
\sum_{L=0}^{L_c}
V_{ab}^{(L),r}\,
A_{\mu\nu}^{(L),(a,b)}.
\label{eq:Vee2index}
\end{equation}
Here $L_c$ is taken large enough that the omitted tail is negligible.
Equation~(\ref{eq:Vee2index}) is the central structural result for the combined
radial-angular basis: although the angular basis varies from shell to shell, the
electron-electron interaction still enters only as a two-index object on the
one-particle basis labels $(a,\mu)$.  The expensive four-index Coulomb tensor is never
formed.   

In second-quantized form, the Hamiltonian is
\begin{equation}
\hat H
=
\sum_{a\mu,b\nu}
h_{a\mu,b\nu}\,
c_{a\mu}^\dagger c_{b\nu}
+
\frac{1}{2}
\sum_{a\mu,b\nu}
V_{a\mu,b\nu}\,
\hat n_{a\mu}\hat n_{b\nu},
\label{eq:Hsecondq}
\end{equation}
with $\hat n_{a\mu}=c_{a\mu}^\dagger c_{a\mu}$, and normal ordering is inserted for
$a=b$ in the interaction to remove self-interaction.
Thus the one-electron part is exact Galerkin, while the two-electron part retains the
density-density form that motivated gausslets from the beginning.  The resulting basis
is flexible enough to adapt its angular resolution shell by shell, but still compact
enough that Hartree--Fock and later tensor-network treatments remain practical. 

\subsection{Simple atomic benchmarks}
\label{sec:simplebench}

Before turning to matrix-product-state calculations, we test the coupled radial-angular
Hamiltonian in settings where the many-body solver is not the limiting factor. We therefore
examine first-row Hartree--Fock errors and the correlated two-electron He problem by exact
diagonalization.

\begin{figure}
  \centering
  \includegraphics[width=0.8\columnwidth]{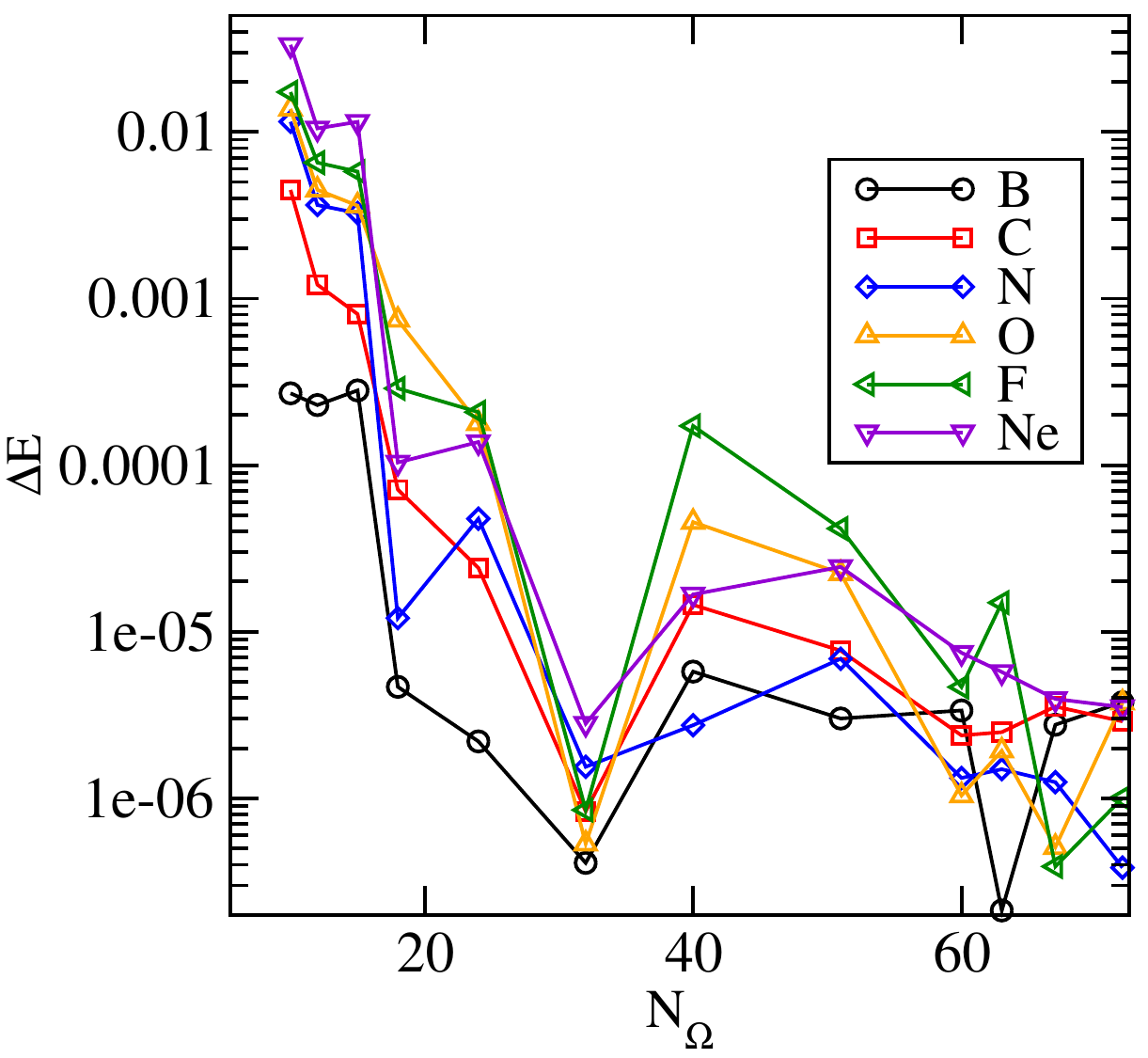}
\caption{Hartree--Fock total-energy errors in Hartrees for first-row atoms from B through Ne, using the
  shell-dependent angular gausslet basis coupled to the radial gausslet construction. The error
  $|E-E_{\rm ref}|$ is plotted against the number $N_\Omega$ of angular centers per shell.
  The $N_\Omega=10$ point is the first order at which the exact $L=1$ sector is injected.
  The especially sharp improvement seen for $N_\Omega=32$ is not accidental:
  this is the first order here with exact injection through $L=3$, and it also uses the
  unusually regular 32-point sphere set discussed in Sec.~\ref{sec:spherepoints}.
  Beyond that threshold, increasing $N_\Omega$ further improves the accuracy of the atom-level
  calculation, showing that practical convergence is controlled not only by exact low-$L$
  injection but also by the quality of the localized angular representation and the two-index
  interaction approximation.}
    \label{fig:firstrowhf}
\end{figure}

Figure~\ref{fig:firstrowhf} shows that the coupled radial-angular construction gives a
  systematic and practically useful Hartree--Fock description across the heavier first-row
  atoms. For B, C, O, and F, microhartree accuracy is already reached by $N_\Omega=32$, while N
  requires somewhat larger angular orders, and Ne reaches a few-microhartree accuracy already at
  $N_\Omega=32$. The pronounced dip seen for $N_\Omega=32$ is also physically
  reasonable: this is the first order with exact injection through $L=3$, and it coincides with
  the unusually regular 32-point set discussed in Sec.~\ref{sec:spherepoints}. Thus the exact low-$L$ injection threshold is not by itself the full convergence
  criterion for atomic calculations: once the relevant low angular channels are present
  exactly, further improvement still comes from refining the localized angular representation
  used in the shell-dependent Hamiltonian and in the two-index treatment of the electron-electron
  interaction. This is the atomic analogue of the purely angular benchmarks in the previous
section. There, Fig.~\ref{fig:KE32}, the low-$\ell$ Coulomb diagnostics collected in Appendix~\ref{app:diagnostics}, and the spherium test
  established the one-body angular Hamiltonian, the low-$\ell$ interaction approximation, and
  a direct purely angular two-electron benchmark; here the same construction is shown to
  remain effective when coupled to the radial gausslet basis in realistic atoms.

\begin{figure}[t]
\centering
\includegraphics[width=0.7\columnwidth]{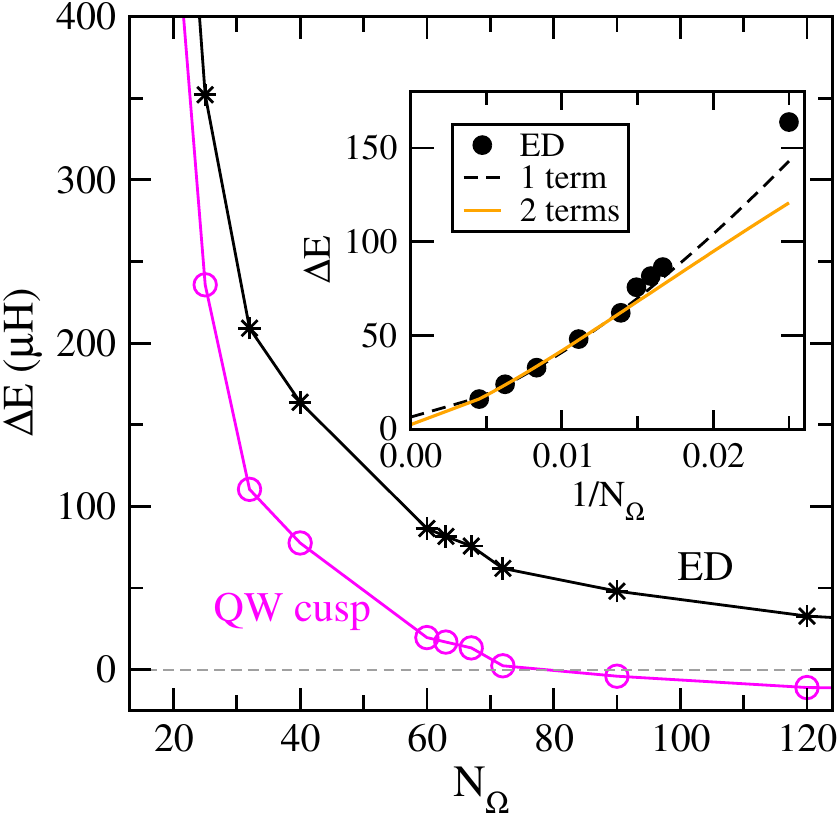}
\caption{Ground-state energy error for the He atom from two-electron exact diagonalization in
  the shell-dependent angular gausslet basis, plotted versus the number $N_\Omega$ of angular
  centers per shell. The black curve shows the raw ED error $E-E_{\mathrm{exact}}$, while the
  orange curve shows the Qiu-White double-occupancy cusp correction applied a posteriori using
  the universal parameters $e_0=-0.005078$ and $\alpha=0.79$. Both curves are shown relative to
  the exact He ground-state energy, with errors reported in microhartree. The raw ED result
  converges systematically with increasing $N_\Omega$, and the cusp correction substantially
  reduces the remaining basis error, bringing the best points down to the low-microhartree
  level. Inset: raw ED errors plotted against $1/N_\Omega$, together with one-term and two-term
  fits over the higher-$N_\Omega$ regime.}
  \label{fig:HeED}
\end{figure}

Figure~\ref{fig:HeED} extends the atomic benchmarks beyond Hartree--Fock to a correlated two-
electron calculation. Because He has one electron of each spin, it is a natural setting for
exact diagonalization with the present radial-plus-angular Hamiltonian, and it also provides
a clean test of the two-index interaction approximation in a genuinely correlated atom. The
raw ED energies converge steadily as the angular order is increased, reaching tens of
microhartree accuracy at the largest practical $N_\Omega$ values examined here. The Qiu-White
double-occupancy cusp correction, evaluated a posteriori from the ED wavefunction, further
reduces the remaining error and gives low-microhartree accuracy at the best points. The
correction is clearly helpful, although not strictly monotone in $N_\Omega$.
The raw He ED sequence is also useful methodologically. The inset suggests that the
higher-$N_\Omega$ raw errors are broadly consistent with a leading angular-tail form
$N_\Omega^{-3/2}$, together with visible next-term curvature. This is in line with the classic
partial-wave convergence analysis of atomic correlation energies,\cite{Schwartz62,KutzelniggMorgan92,BromleyMitroy07}
where truncation at $\ell_{\max}$ gives a leading $(\ell_{\max}+1)^{-3}$ error; since the number of angular functions scales roughly as
$N_\Omega \sim \ell_{\max}^2$, $N_\Omega^{-3/2}$ is a natural variable to test here. 

\section{Matrix product state methods and correlated Be results}
\label{sec:mpsmethods}
\label{sec:mpsbe}

In conventional ab initio quantum-chemical DMRG, active spaces treated directly
by DMRG with a few dozen orbitals are routine, spaces in the $\sim 50$--80
orbital range are high-end, and calculations beyond $\sim 100$ orbitals, while
possible, remain exceptional and usually rely on especially favorable orbital
structure or unusually optimized parallel
implementations.\cite{WoutersVanNeck2014,SteinReiher2016,BaiardiReiher2020,Brabec2021,ZhaiChan2021}

By that standard, a naive direct DMRG treatment of an angular-gausslet basis,
without isolating an active space and while directly treating dynamic
correlation, would appear out of reach. The crucial point, however, is that a
main obstacle in standard QC-DMRG is the generic four-index
two-electron interaction. In the angular-gausslet basis, the integral-diagonal
approximation compresses this part of the Hamiltonian drastically, making it
much more favorable for DMRG. Similar diagonal or partially diagonal
interactions have been exploited in earlier gausslet and sliced-basis works on
hydrogen chains, where thousands of basis functions were
treated.\cite{StoudenmireWhite2017,WhiteStoudenmire2019,SawayaWhite2022,QiuWhite2021}

First-row atoms in high-resolution localized bases are much harder than
hydrogen chains for DMRG because increasing nuclear charge $Z$ creates much
sharper near-nucleus length scales and a stronger core/valence structure, while
the atomic problem also retains intrinsically three-dimensional geometry and angular
couplings. By contrast, making a hydrogen chain longer mainly extends a
quasi-1D geometry that is intrinsically favorable for DMRG. Here we demonstrate
that a direct DMRG treatment, including dynamic correlation well below chemical accuracy, 
is feasible in the case of the Be
atom.\cite{StoudenmireWhite2017,WhiteStoudenmire2019,QiuWhite2021,WhiteLindsey2023}
This does not mean that this route is necessarily preferred over, say, separate treatments
of active and virtual spaces. But it serves as a milestone for considering other possible 
variations which may be more efficient.  To facilitate this path, the main property is the diagonal 
interaction, but in addition, we develop techniques which speed up the DMRG calculations and facilitate 
extrapolations.

In the Be calculations below, the radial part of the basis is held fixed at a deliberately
conservative choice: the basis extends to $R_{\max}=30$ bohr and uses a fine radial scale parameter 
$s=0.2$, with 41 radial functions, so the residual radial incompleteness should already be small on
the scale of the angular effects of interest. What is varied is only the
angular resolution. We also impose a deliberately
simple shell policy, using the same number $N_\Omega$ of angular functions on
every radial shell, even though such uniform angular resolution is not really
needed either near the nucleus or in the far
tail. This makes the sequence easy to interpret and gives a clear range of many-body problem
sizes: the corresponding one-particle bases contain $410$, $615$, $984$, $1312$, and $1968$
basis functions for $N_\Omega=10$, $15$, $24$, $32$, and $48$, respectively.

\subsection{Matrix product operator (MPO) representation of the atomic Hamiltonian}

The starting point is the second-quantized Hamiltonian of Eq.~(\ref{eq:Hsecondq}), with exact
one-electron matrix elements and a two-index density-density interaction. We use separate MPOs for the
interaction term, for pieces of the one-electron term, and for a small remaining diagonal
term. An important
feature of the atomic angular-gausslet Hamiltonian is that the
interaction is not the larger MPO! The two-index interaction compresses very well, so the
resulting interaction MPO is smaller than the one-electron MPO. For the Be atom with
$N_\Omega=48$, for example, representative maximum MPO bond dimension was about $300$ for the 
interaction MPO. Even when split into four pieces with an extra compression step (see below), each 
single-particle MPO had 
dimension about $460$. These four pieces correspond to the spin-up and spin-down hopping terms,
split additionally by whether a $c^\dagger$ or a $c$ site appears first in the chain.
Only one of the MPOs in this last split needs to be created and worked with; the other comes from Hermitian 
conjugation when needed. This saves a factor of two in the main storage term for DMRG. (A small diagonal
term MPO is treated separately.) Without this splitting, a single one-particle MPO would have a size well over $1000$. 

The extra MPO size reduction step is to keep track of the exact truncations in turning $H_1$ into
a truncated MPO, and use them to form a residual matrix, $\Delta H_1 = H_1 - H_{1,\mathrm{repr}}$,
where the MPO exactly corresponds to $H_{1,\mathrm{repr}}$.
The residual $\Delta H_1$ is restored
after the DMRG run as a first-order correction through the measured total one-particle 
density matrix $\gamma$:
\[
E_{H_1}^{\mathrm{full}} = \mathrm{Tr}(\gamma H_1)
= \mathrm{Tr}(\gamma H_{1,\mathrm{repr}})
+ \mathrm{Tr}(\gamma \Delta H_1).
\]
This split allows much more aggressive truncation with negligible final $H_1$ errors.

\subsection{Initial-state preparation}

For these atomic Hamiltonians, initial-state preparation is not a minor convenience. The
one-particle Hamiltonian spans a wide range of scales, it is neither 1-D coupled nor 
diagonally dominated, and so the number of sweeps needed for
useful convergence depends strongly on the starting state. A simple product state is poor, a
Hartree--Fock determinant is acceptable, a small-space correlated starting state is much better,
and a state lifted from a neighboring basis in an angular sequence is better still. Our
implementation uses the latter three levels in sequence, with Givens rotations providing the
link between them.

For the first basis size in a sequence, we begin with a Hartree--Fock calculation followed by second
order perturbation theory (MP2), construct the corresponding approximate natural orbitals (NOs), 
and select from them a small most-occupied set--the small basis. 
We project the Hamiltonian into the small basis, and run a short DMRG calculation there, starting
with the projected HF state.
The resulting correlated small-space MPS is
then rotated and embedded back into the full basis as an MPS, using Fermionic 
Givens rotations. Call each basis function a site: each Givens rotation is a 
two-site gate, which performs a unitary single 
particle rotation for those two sites, expressed in many-particle form\cite{Fishman}.
It has only one degree of freedom, a rotation angle between those sites. Here is a simple recipe
for determining the angles: select the first NO, and start on the last site.  Find the rotation 
that makes the last site orthogonal to the first NO. Then find the rotation of the adjacent 
link that makes the second to last site orthogonal to the first NO, and continue, eventually
leaving the first NO on the first site. Continue with all the rest of the small space. 
These gates would take an MPS in the full space and turn it into a small space (throwing away
the leftover sites at the end). Reversing the gates transforms the small MPS into the large. This construction leaves an arbitrary minus sign for each NO, which can be tracked and removed with a one-site gate.

In the current calculations, we choose a small space size of $4Z$ orbitals, but 
we have experimented with significantly larger small spaces.
Its Hamiltonian is in full four-index form, but its small size makes the DMRG calculation very quick.
After this short correlated starting calculation, the resulting state is transformed to the full basis.

This small correlated starting
state gives a much better first full sweep than a pure Hartree--Fock state while remaining much
cheaper than a production run on the full basis, because the small problem is solved only once
during initialization. Later basis sizes in the angular sequence then start from states lifted from
the previous size rather than from scratch.

\subsection{Transfer along an angular sequence}

One of the main practical advantages of the shell-dependent angular construction is that calculations
on bases with nearby values of $N_\Omega$ are closely related. 
We therefore do not treat each value of $N_\Omega$ as
an isolated problem: we start small and work our way up. The converged DMRG MPS from one $N_\Omega$
becomes the initial state of the next higher $N_\Omega$ via a set of Givens rotations. 
The gate sets are reduced in size, since the DMRG path keeps the angular functions from a
given radial shell together before moving to the next shell. The Givens rotations stay within shells,
since the radial basis is identical.

The construction of these gates is subtle because the smaller shell basis is not exactly embedded
in the larger one. The raw source-target shell overlap is therefore not itself a norm-preserving
many-body basis change. We instead take the polar decomposition of that within-shell overlap and
use its closest norm-preserving factor as the actual one-particle map. Since the same angular
profile overlap is reused on every radial shell, this gives one common shell map for the whole
adjacent $N_\Omega \rightarrow N_\Omega'$ step. That shell map is then completed to a full
orthogonal transformation on the target shell and factorized into ordinary two-site Givens
rotations, which are applied shell by shell to rotate the source MPS into the larger basis.

To quantify the quality of the transfer itself, we introduce an occupancy-weighted leakage
\begin{equation}
\xi = \frac{\Tr\!\left[\gamma\left(I-OO^T\right)\right]}{\Tr(\gamma)},
\end{equation}
where $O$ is the full one-particle source-to-target overlap matrix and
$\gamma$ is the one-particle density matrix of the source state in the
source basis. Thus $\xi=0$ for an exact embedding, while overlap defects
in directions carrying little one-particle weight contribute only weakly.

This transfer machinery provides an excellent starting state for the next DMRG, which gets better
with larger $N_\Omega$.
As an added benefit, it sets up natural sequences of calculations and extrapolations in $N_\Omega$, at fixed
radial basis.

\subsection{Variance extrapolation and ESVE}

In model-system DMRG it is common to extrapolate in the discarded weight or truncation
error, but that route is less useful here. It requires a sequence of runs at different bond
dimensions, whereas the transfer methods above are designed to avoid doing that at every
basis size, and for the present three-dimensional gausslet site traversals we find the
energy versus truncation-error relation less regular than in more one-dimensional settings.

For a normalized many-body state $\ket{\Psi}$ and a fixed Hamiltonian $H$, the exact variance
\begin{equation}
\sigma_H^2 = \avg{H^2} - \avg{H}^2
\end{equation}
vanishes for an exact eigenstate, so it is a natural extrapolation variable.
The difficulty is that for a large Hamiltonian MPO, evaluating $\sigma_H^2$
exactly for a DMRG state is expensive and memory-intensive.

A cheaper route is to sample configurations $x$ from the probability distribution $p(x)=|\braket{x}{\Psi}|^2$, where $x$ is a site-occupation product state. For each sample we form the local energy
\begin{equation}
E_{\mathrm{loc}}(x) = \frac{\bra{x}H\ket{\Psi}}{\braket{x}{\Psi}}.
\end{equation}
This use of a local-energy estimator is standard in variational Monte Carlo,\cite{Foulkes2001QMC} and Silvester, Carleo, and White\cite{Silvester} showed how to use it effectively for variance extrapolation with MPS wavefunctions.

With that sampling distribution one has
\begin{equation}
\sigma_H^2 = \langle (E_{\mathrm{loc}}(x)-E)^2\rangle_p + \Delta_s
\end{equation}
where $\Delta_s$ comes from configurations with $p(x)=0$ but $\bra{x}H\ket{\Psi}\ne 0$. In practice this correction is negligible. The raw sampled variance, however, is very noisy, because the local-energy distribution of an MPS has broad high-energy tails.\cite{Silvester} Following Ref.~\cite{Silvester}, we therefore use a bounded sampled variance: after sorting the sampled local energies by their distance from the sampled mean $E$, we keep only a central fraction $c$ of the samples and define $\sigma_c^2$ from that retained set. Here we use $c=0.85$ and $0.9$, denoting the resulting bounded sampled variances by $\sigma_{85}^2$ and $\sigma_{90}^2$. Although this trimming introduces a bias, that bias decreases smoothly as the state approaches an eigenstate, so it can be absorbed into the extrapolation. We will refer to this procedure as sampled variance extrapolation (SVE).
In the Be calculations below, we explored both $c=0.9$ and $c=0.85$ and examined the statistical
errors in the variance surrogate as a function of sample count. For the largest systems we judged
2000 samples to be sufficient, with statistical errors below about $10\%$, which is adequate for
the variance extrapolations used here.

Ordinary SVE is defined for one fixed Hamiltonian and basis. Here we use the same idea along the angular sequence, where nearby basis sizes are connected by the shell-local transfer maps described above. Let $\ket{\Psi(N_s)}$ be a DMRG state obtained in a source basis of angular size $N_s$, and let $\ket{\widetilde\Psi(N_s\!\to\! N_t)}$ denote the transferred state in a larger target basis $N_t\ge N_s$. For that embedded state we evaluate the target-Hamiltonian energy
\begin{equation}
E(N_s,N_t) =
\bra{\widetilde\Psi(N_s\!\to\! N_t)} H(N_t)
\ket{\widetilde\Psi(N_s\!\to\! N_t)}
\end{equation}
which differs slightly from the source-basis energy because of the transfer and because the IDA is re-evaluated on the target system. We then sample the same transferred state in the target basis to obtain a bounded sampled variance there. Repeating this for several source sizes $N_s$ gives a family of energy-variance points for one fixed target $N_t$, from which we can extrapolate a better estimate of the target-basis energy.

This is useful because the target calculation only requires energy evaluations and sampled local energies for the transferred states; it does not require a full DMRG optimization at the target size. In our tests this reduced the memory requirement substantially and was also much faster than continuing the full DMRG calculation. We refer to this construction as embedded sampled variance extrapolation (ESVE).

ESVE involves two errors at once: residual many-body error in the source DMRG state and basis incompleteness of the source angular space. The transferred state, viewed in the target basis, reflects both. If those two errors affect the energy-variance relation in a similar way, the extrapolation can remove much of both. Of course, it is still an extrapolation, so it should not be pushed too far: one expects it to degrade once $N_t$ becomes much larger than the available source sizes.

For the present Hamiltonian there is also a practical computational advantage. In the ESVE
sampling stage, the sampled local energy is evaluated directly rather than through a sampled MPO
contraction. For the interaction term, the two-index density-density form gives
$V_{\mathrm{loc}}(x)$ immediately as an occupation-pattern sum for the sampled configuration
$x$. For the one-electron term, $H_{1,\mathrm{loc}}(x)$ is evaluated by a direct single-hop
expression using left and right MPS edge contractions and summing over all allowed one-particle
moves from $x$. This removes the sampled MPO-bond overhead from the dominant sampled-energy
loop. If the MPS bond dimension is $m$ and the sampled-MPO bond dimension is $k$, then a sampled
MPO contraction costs roughly $m^2k + mk^2$ per site per sample, whereas the direct
one-electron evaluation is instead controlled by the number of occupied particles $N_e$ and
scales roughly as $N_e m^2$ per site per sample. Note that $N_e \ll k$.

Because the transferred states are already very accurate, the one-electron expectation value
$\avg{H_1}$ changes only weakly when the state is embedded into a larger target basis. In the Be
transfers studied here, holding $\avg{H_1}$ fixed at its source-system value and updating only
$\avg{V}$ in the larger target already gives errors of only about $0.008$ to $0.015$ mH for
typical useful transfers, and as little as $0.0029$ mH for the $48\to 80$ case. Since the sampling is very fast, even this removal of a straightforward evaluation of $\langle H_1 \rangle $ in the target system is a noticeable improvement in ESVE cost. 

\subsection{Correlated Be results}
\label{sec:bedmrg}

With these methods in place, we now turn to Be. It is the first atom for which the
angular gausslet Hamiltonian is large enough that exact diagonalization is no
longer the practical method of choice, so it is a natural first benchmark system for both the
basis convergence beyond two particles and the DMRG techniques introduced above. We first
examine bond-dimension convergence, then transfer quality, then fixed-target ESVE families,
and finally the target-size extrapolation. The calculations that follow were done almost
entirely on one modest desktop, a 64 GB M4 Mac mini, within about a week's total time, i.e.\
several overnight runs. Clearly one could do much more on even a modest cluster, but this
constraint also serves as a useful reference point.

\subsubsection{DMRG convergence in an angular basis}
Before detailing the DMRG lifting and extrapolation techniques, we address
simple DMRG convergence with the bond dimension. The basis sequences were all
based on identical fairly fine and extensive radial gausslet bases with $s=0.2$
and $R_{\rm max}=30$, with 41 radial functions. We first consider the largest
system in our sequence, $N_s=48$, which had a total basis size of
$41\times48=1968$. The path through the basis started out at the innermost
shell, and proceeded through the sites shell by shell, with the ordering within
a shell taking a Fiedler order based on the off-diagonal elements of the
single-particle Hamiltonian matrix\cite{BarczaLegezaMartiReiher2011}. The entire wavefunction transfer DMRG
sequence was carried out at very high accuracy, specifically a truncation error
of about $10^{-11}$ at bond dimension 600. Starting with this final ground
state at size $48$, we then studied how large a bond dimension was needed for a
given accuracy, by
truncating the state to a specific bond dimension, and measuring the resulting
energy. Only modest bond dimensions were needed for excellent accuracy: at bond
dimension 100, we were at chemical accuracy, 1.6 mH. At 200, the error was 0.24
mH, at 300, 0.05 mH, and at 400, 0.01 mH. We believe this excellent convergence
with bond dimension reflects two factors: first, the modest number of
electrons, and second, the sharp locality of the radial basis.

\begin{figure}[t]
\centering
\includegraphics[width=0.7\columnwidth]{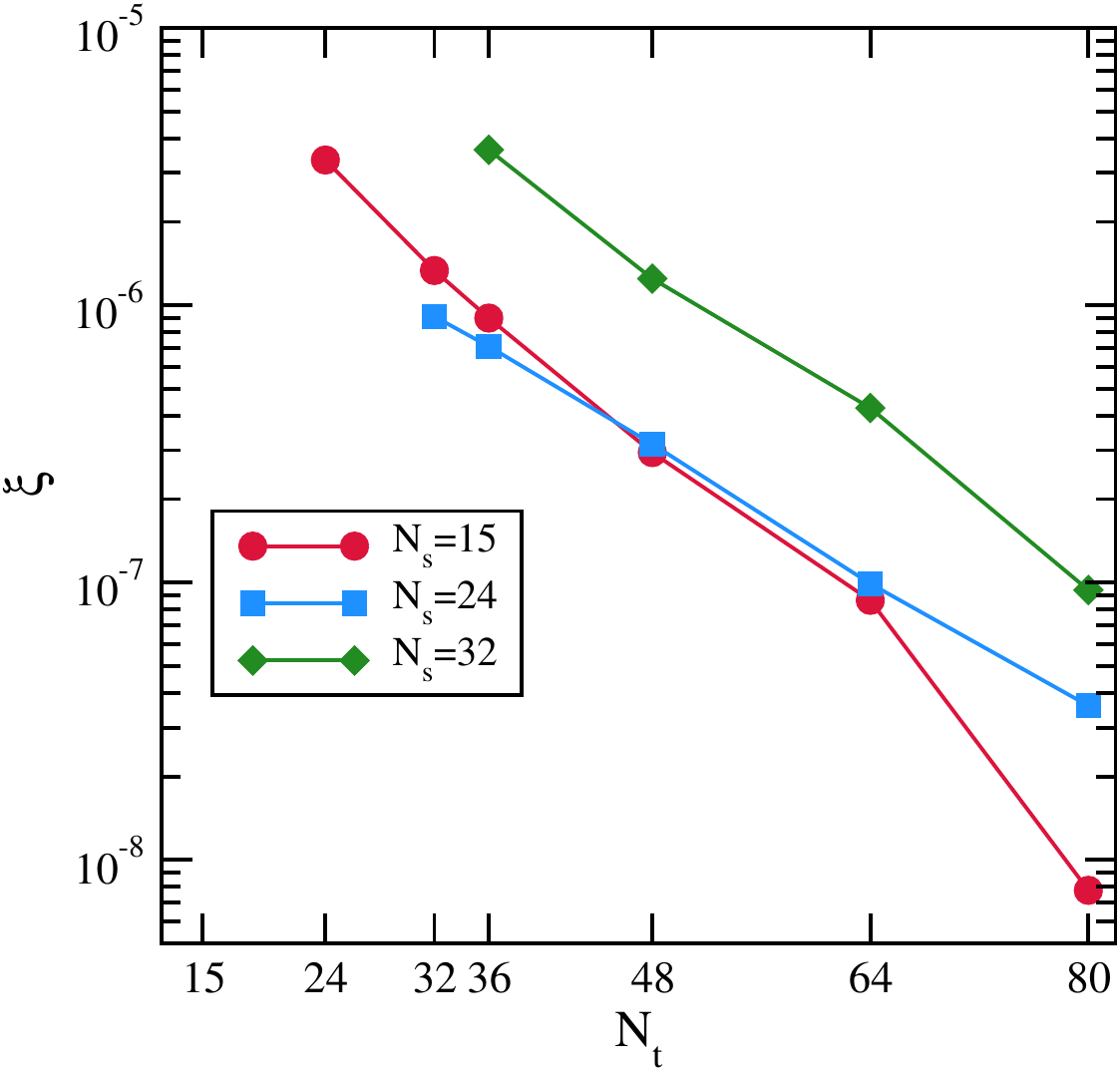}
\caption{Transfer diagnostic $\xi$ versus target angular size $N_t$ for fixed
source sizes $N_s=15,24,32$.  The monotonic decrease of $\xi$ with
increasing $N_t$ shows that the transfer becomes more faithful as the target
angular space is enlarged.}
\label{fig:transfer}
\end{figure}

\subsubsection{Wavefunction transfer}
Our first sequence had angular sizes of
$N_s = 10, 15, 24, 32, 48$. Subsequently, we added $N_s=36$, transferring the starting MPS for the DMRG from the $N_s=32$ system. The reason for considering both 32 and 36, nearby points, was that 32 in the HF studies was exceptionally accurate, unlike the other points, so that it might not be useful for an extrapolation sequence. From the initial sequence it was immediately clear that $N_s=10$ was an outlier, too small for any extrapolations in $N_s$. Across the sequence the corrected energies improve monotonically,
showing that the angular basis continues to give systematic gains in the correlated regime. 

In Figure~\ref{fig:transfer} we quantify the quality of the source-to-target transfer using the occupied-weighted leakage $\xi$. The different curves are for several fixed source states as the target angular size $N_t$ is increased. In a typical sequence the next $N_s$ would serve as the target and is most relevant, but here we see that $\xi$ decreases strongly with increasing $N_t$, indicating that the larger target spaces provide more accurate embeddings of the same source state. This supports the use of transferred states both as practical starting states for DMRG and as the basis for the fixed-target ESVE analysis. Given the excellent starting state, in practice only a few DMRG cleanup sweeps are needed after transfer.

The transfer is highly accurate, but the diagonal IDA form of the interaction energy is approximate, so the interaction energy of a fixed transferred state does not vary completely smoothly with target size. To estimate this IDA error, we transfer each source state to our largest target, $N_t=80$, and take the resulting $\langle V\rangle$ as the best available interaction reference value for that state. This gives estimated IDA errors in $\langle V\rangle$ of 0.0460, 0.0030, 0.0387, 0.0213, and 0.0007 mH for $N_s=15,24,32,36,$ and 48, respectively. Thus the IDA error in $\langle V\rangle$ is always below 0.05 mH, although it is not strictly monotonic along the sequence.

\subsubsection{Variance extrapolations: SVE and ESVE}
\begin{figure}[t]
\centering
\includegraphics[width=0.82\columnwidth]{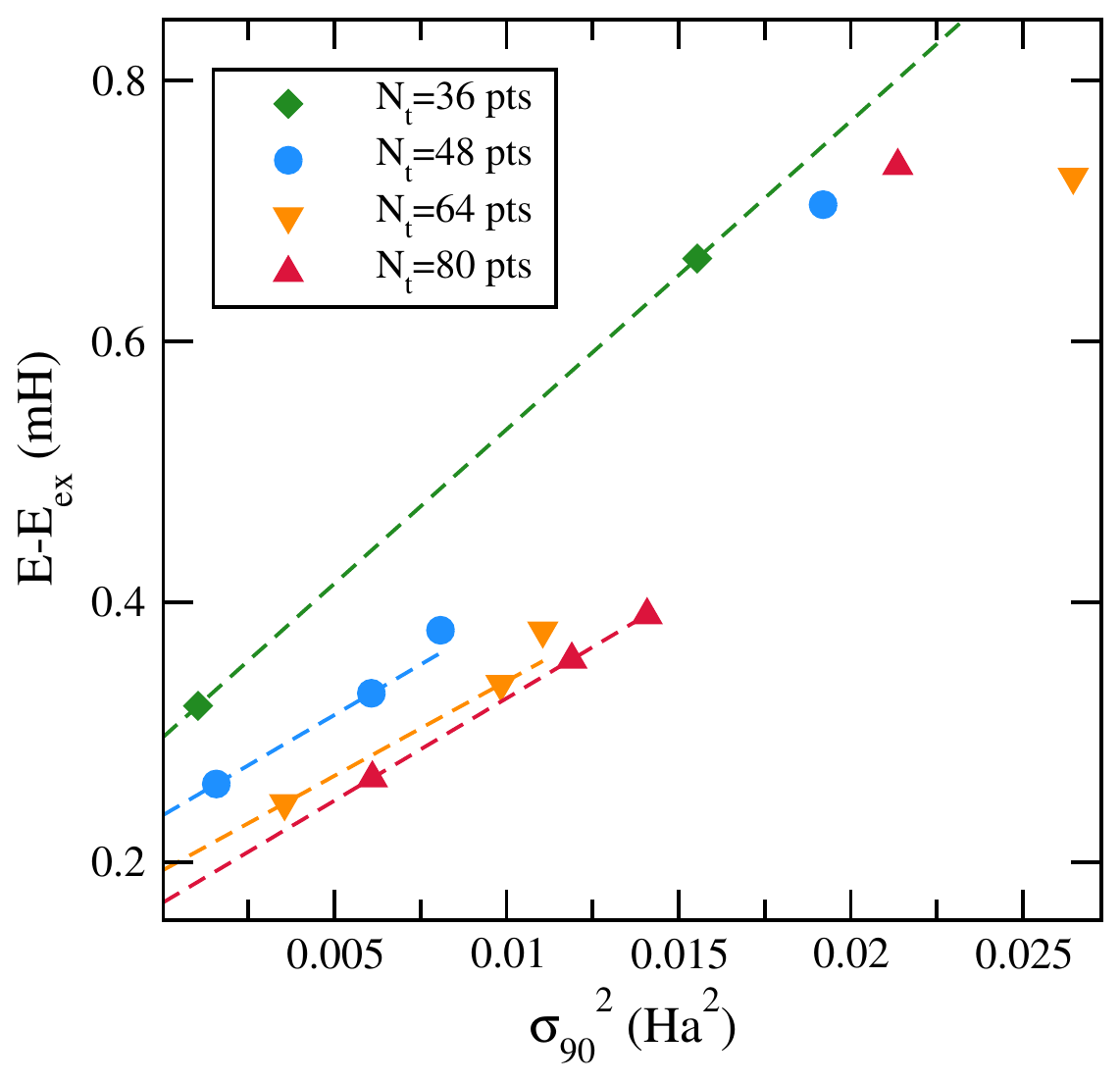}
\caption{Variance extrapolation to reach larger systems using ESVE for the Be atom. Extrapolations for $N_t=36$, $48$, $64$, and $80$ are plotted as
$E-E_{\mathrm{ex}}$ versus the bounded sampled-variance surrogate $\sigma_{90}^2$. Each symbol represents a smaller system $N_s$ transferred to the size $N_t$. The dashed lines are local line fits through the lowest-variance end of each family. The families are close to linear over their low-variance ends, with larger deviations only when $N_s \ll N_t$. }
\label{fig:BeESVE}
\end{figure}

How close can we come to the complete basis set fully correlated Be energy ($E_{\rm ex}=-14.6673561335$ Ha\cite{Sims2014Be}), using large basis sets and extrapolation in the angular basis size, but no cusp corrections? Because relatively small bond dimension was needed for highly accurate DMRG, ordinary SVE to improve the ground state energy for a particular system was not particularly important. However, ESVE allows us to reach larger systems than could be studied with DMRG, and we extended the target-size sequence to $N_t=64$ and 80. First we consider how well behaved the variance extrapolations are. Figure~\ref{fig:BeESVE} summarizes the fixed-target variance data on one common energy scale. The behavior is approximately linear, allowing reasonable extrapolations to zero variance. ESVE does not provide variance extrapolation to $N_t \to \infty$; $N_t$ is always finite. A separate more conventional extrapolation is needed for $N_t \to \infty$, but the added ESVE points assist in making that extrapolation accurate. Note that because the radial discretization is held
fixed at $s=0.2$, the $N_t\to\infty$ limit of this angular extrapolation is not expected to
coincide exactly with the nonrelativistic Be energy $E_{\mathrm{ex}}$.

\subsubsection{Final target-size extrapolation}
\label{sec:befinal}

\begin{figure}[t]
\centering
\includegraphics[width=0.82\columnwidth]{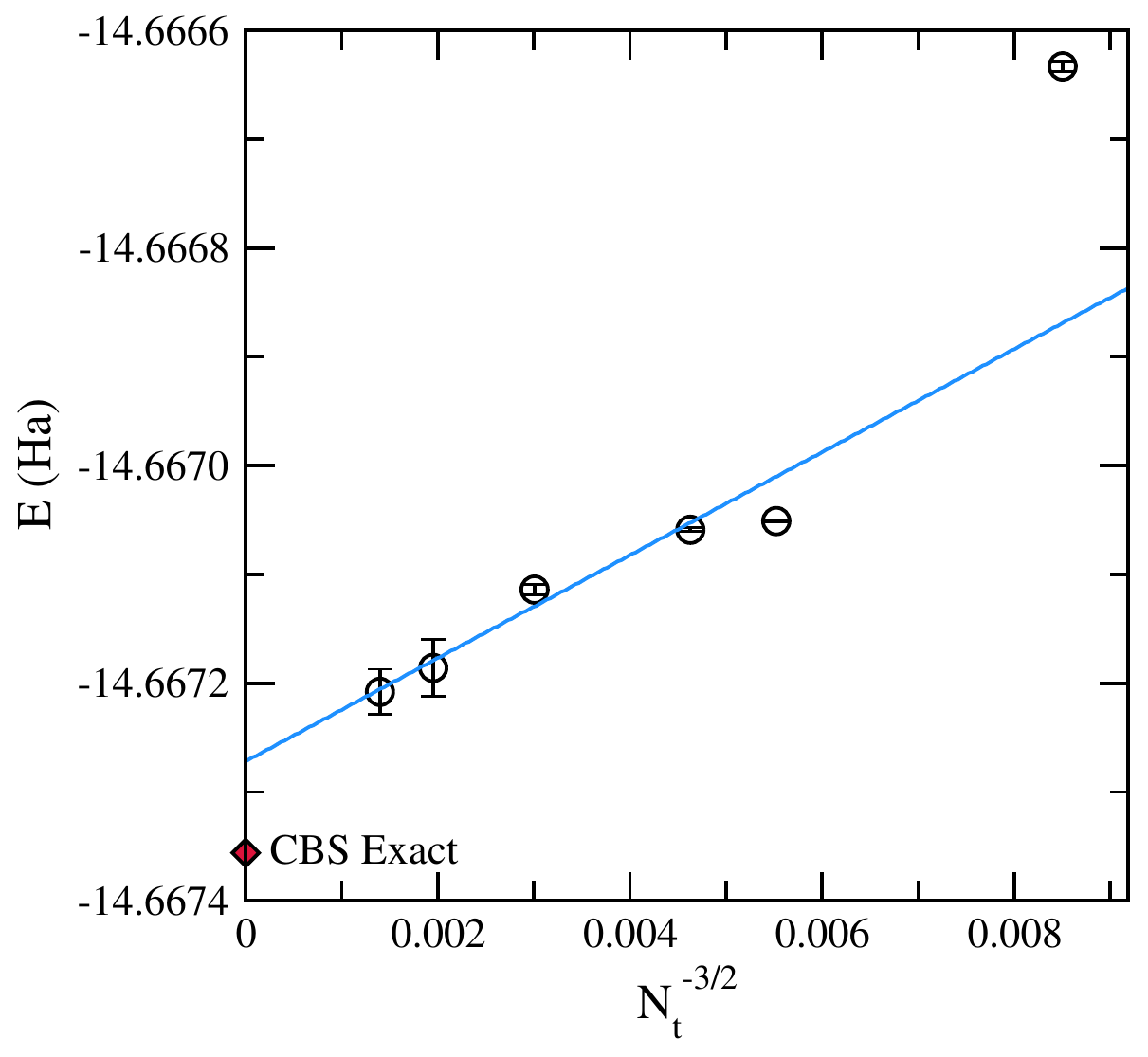}
\caption{Final target-size extrapolation for Be. The corrected zero-variance target estimates are plotted versus $N_t^{-3/2}$. All circles are target estimates, while the solid line is a one-term $N_t^{-3/2}$ fit using only the tail points $N_t=36,48,64,$ and 80. The lower points $N_t=24$ and 32 are shown for context but are not included in the fit. The red diamond at $N_t^{-3/2}=0$ marks the complete-basis-set exact result. The fitted fixed-radial angular limit is $-14.66724$ Ha. The difference between these reflects the fixed radial basis size.}
\label{fig:BeFinalExtrap}
\end{figure}

From the ESVE analysis, 
we extract corrected zero-variance target estimates for $N_t=36$, 48, 64, and 80, and then fit the remaining angular convergence as a function of target size. Motivated by the He data and by the corresponding partial-wave asymptotics just noted,\cite{Schwartz62,KutzelniggMorgan92,BromleyMitroy07} we treat $N_t^{-3/2}$ as the leading target-size variable. Figure~\ref{fig:BeFinalExtrap} shows the resulting target-size sequence. The points $N_t=36,48,64,80$ form a clean near-linear sequence in $N_t^{-3/2}$, while the lower points $N_t=24$ and 32 are shown for context but excluded from the fit. The corresponding fit gives an extrapolated fixed-radial limit of
\[
E_\infty = -14.66724 \ \mathrm{Ha}.
\]
This is about 0.12 mH above the exact nonrelativistic Be energy, which is the expected behavior here: the extrapolation removes the residual angular error while the radial discretization remains fixed at $s=0.2$. Thus the extrapolated limit should approach the fixed-radial Be value, not the exact continuum value.

\section{Discussion and outlook}
\label{sec:discussion}

We have introduced angular generalized gausslets and combined them with radial gausslets to form an atom-centered basis in which the electron-electron interaction takes a two-index integral-diagonal form. This extends the radial-gausslet construction, where the angular dependence was still kept in spherical harmonics and the interaction was only partially diagonalized. The new element is a localized angular basis on the sphere built from spherical Gaussians with exact injection of a low-$\ell$ spherical-harmonic subspace.

The numerical tests show that angular gausslets with the IDA integral diagonal approximation work very well, giving accuracy well below chemical accuracy and systematically converging to the complete basis set limit.  The kinetic spectrum, low-$\ell$ Coulomb tests, and spherium benchmark validate the angular part on its own. First-row Hartree--Fock calculations and exact diagonalization for He show that the same construction continues to work well when coupled to radial gausslets in real atomic calculations.

The Be calculations show that direct DMRG is possible in this atomic large-basis setting, up to about 2000 basis functions, with both static and dynamic correlation treated within the DMRG itself. This goes beyond the earlier gausslet-based demonstrations on hydrogen chains, where the geometry was much more favorable to DMRG. Because nearby values of $N_\Omega$ are closely related, converged states can be transferred from one angular size to the next rather than rebuilt from scratch. Together with compact MPOs and a correlated small-space starting calculation, this makes direct DMRG feasible for the present atom-centered Hamiltonian. The same transfer machinery also makes ESVE possible on larger target spaces and provides a practical route to controlled angular extrapolations beyond the largest direct calculations. In the present fixed-radial, no-cusp setting, the resulting Be energies already reach the few-tenths-of-a-millihartree accuracy level, which is enough to show that the method is quantitatively useful and not only qualitatively feasible.

Several extensions appear promising. Even within the present atomic setting, the current construction is only one very simple choice. One could vary the angular resolution more aggressively with radius, or contract a core region into local block natural orbitals, giving up the pure IDA structure there in exchange for a more compact treatment of the most tightly bound region. Another interesting question is whether one can introduce an active/virtual separation in a way that preserves much of the diagonal structure, rather than immediately giving it up everywhere once such a separation is made. 

More ambitiously, one can consider embedding angular gausslet atomic regions inside a larger Cartesian-gausslet description, so that specialized atom-centered resolution is combined with a more general surrounding basis. On the many-body side, the transfer and ESVE ideas should remain useful for larger atoms and for future localized molecular calculations. More broadly, the present work shows that the generalized gausslet construction is useful and accurate enough to support direct localized bases in higher dimensions. In particular, it suggests that genuinely three-dimensional generalized-gausslet constructions may be possible, and such direct 3D bases would likely be the most efficient route in terms of achieving a given accuracy with the fewest basis functions.

The basis construction and Hamiltonian generation were performed with the open source \texttt{GaussletBases.jl} software library at \url{https://github.com/srwhite59/GaussletBases.jl}.


\acknowledgments
I thank Sandeep Sharma for helpful discussions. This work was supported by the U.S. NSF under Grant
DMR-2412638. A Julia implementation of the angular gausslet constructions described here is available in the public repository https://github.com/srwhite59/GaussletBases.jl. The author has no conflicts to disclose.

\noindent{\bf Author Contributions}
{\it Steven R. White}: Conceptualization; Methodology; Software; Validation; Formal analysis; Investigation; Data curation; Visualization;
Writing: original draft, review and editing; Funding acquisition.

\appendix

\section{Additional angular diagnostics}
\label{app:diagnostics}

This appendix collects the quantitative diagnostics that support the qualitative
claims made in the main text about angular-basis locality, near-rotational
symmetry, and the accuracy of the low-$\ell$ Coulomb diagonal approximation.
It also records the concrete metrics used to choose the default parameter
regime for the angular basis.

\subsection{Angular-basis diagnostics}

For each final angular orbital $\phi_a(\Omega)$, we assign a center $\hat n_a$ by the prototype Gaussian carrying the largest Gaussian-part coefficient,
\begin{equation}
j(a)=\operatorname*{arg\,max}_j |D_{ja}|,
\qquad
\hat n_a \equiv \hat n_{j(a)},
\end{equation}
where $D_{ja}$ are the Gaussian-part coefficients of the contracted form in Eq.~(\ref{eq:Ginjcontracted}), specialized here to the angular basis on the sphere.

The diagnostics used in Table~\ref{tab:angularquality} are built from the weight and first moment,
\begin{equation}
w_a = \int d\Omega\,\phi_a(\Omega),
\qquad
\mathbf m_a = \int d\Omega\, \Omega\,\phi_a(\Omega).
\end{equation}
From these we define the normalized dipole magnitude
\begin{equation}
c_{1,a} \equiv \frac{|\mathbf m_a|}{w_a},
\end{equation}
and the dipole-center shift
\begin{equation}
\delta\gamma_a \equiv
\arccos\!\left(
\hat n_a\cdot \frac{\mathbf m_a}{|\mathbf m_a|}
\right).
\end{equation}
Thus $c_{1,a}$ measures the magnitude of the first moment relative to that of an ideal point mass, while $\delta\gamma_a$ measures the directional shift of that first moment away from the assigned center.

For higher moments we define
\begin{equation}
M_{\ell m}^{(a)} \equiv \int d\Omega\, Y_{\ell m}(\Omega)\,\phi_a(\Omega),
\end{equation}
and then the RMS deviations
\begin{equation}
\epsilon_{\ell,a}
=
\left[
\frac{1}{2\ell+1}
\sum_{m=-\ell}^{\ell}
\left|
\frac{M_{\ell m}^{(a)}}{w_a} - Y_{\ell m}(\hat n_a)
\right|^2
\right]^{1/2}.
\end{equation}
The table reports
\begin{equation}
\epsilon_2^{\max} = \max_a \epsilon_{2,a},
\qquad
\epsilon_3^{\max} = \max_a \epsilon_{3,a},
\end{equation}
together with the spread in weights, the maximum dipole-center shift, and the mean and standard deviation of $c_{1,a}$ over the full set of final angular orbitals.

\begin{table*}
\caption{Locality and low-order moment diagnostics for the final injected angular basis at $\beta=2.0$, for the same three point sets used in Fig.~\ref{fig:worldlines}. Here $N_Y=(L_{\rm inj}+1)^2$ is the dimension of the injected low-$\ell$ spherical-harmonic subspace. The $N=15$ and $N=32$ point sets were obtained by starting from point sets in Sloane's spherical-code tables and then optimizing them, while the $N=51$ set was obtained from a Fibonacci-initialized optimization.}
\label{tab:angularquality}
\begin{ruledtabular}
\begin{tabular}{cccccccc}
$N$ & $L_{\rm inj}$ & $N_Y$ & $w_{\max}/w_{\min}$ &
$\delta\gamma_{\max}$ (deg) & $c_1$ (mean $\pm$ std) &
$\epsilon_2^{\max}$ & $\epsilon_3^{\max}$ \\
\hline
15 & 1 & 4  & 1.293 & 1.245   & $1.0057 \pm 0.0328$ & $3.90\times10^{-2}$ & $1.06\times10^{-1}$ \\
32 & 3 & 16 & 1.012 & $<10^{-5}$ & $0.99997 \pm 0.00213$ & $2.18\times10^{-3}$ & $3.66\times10^{-3}$ \\
51 & 4 & 25 & 1.095 & 0.845   & $1.00003 \pm 0.00207$ & $7.07\times10^{-3}$ & $1.03\times10^{-2}$ \\
\end{tabular}
\end{ruledtabular}
\end{table*}

Several features are evident in Table~\ref{tab:angularquality}. The weight spread
$w_{\max}/w_{\min}$ stays close to 1, especially for $N=32$, consistent with the
unusually regular geometry of that point set discussed in Sec.~\ref{sec:spherepoints}; the dipole-center shifts are
small, and $c_1$ remains very close to the $\delta$-function value 1. The higher-moment
errors $\epsilon_2^{\max}$ and $\epsilon_3^{\max}$ are already modest for $N=15$ and fall to
the $10^{-3}$--$10^{-2}$ range for $N=32$ and $N=51$. These data confirm that the final
angular basis is localized and nearly rotationally symmetric about its assigned centers.
We also used diagnostics like these to choose $\beta=2$ as the default. For $N=32$ and
$N=51$, the best values fell in the range $1.6$--$2.4$, depending on the diagnostic, so
$\beta=2$ is a reasonable compromise.

\subsection{\texorpdfstring{Low-$\ell$}{Low-l} Coulomb IDA diagnostics}

Table~\ref{tab:idaYlm} collects the low-$\ell$ spherical-harmonic-subspace tests of the two-index integral diagonal approximation discussed in Sec.~\ref{sec:tests}. For each basis size $N$, the comparison is carried out in the subspace spanned by all $Y_{\ell m}$ with $\ell \le L_0$, where $L_0 \le L_{\rm inj}$. Here $N_0=(L_0+1)^2$ is the dimension of the tested subspace. The RMS and maximum absolute errors are taken over all four-index Coulomb tensor elements in that subspace, comparing the exact tensor with the tensor reconstructed from the two-index IDA interaction.

\begin{table*}
\caption{Accuracy of the two-index integral diagonal approximation for the Coulomb interaction in low-$\ell$ spherical-harmonic subspaces.}
\label{tab:idaYlm}
\begin{ruledtabular}
\begin{tabular}{cccccc}
$N$ & $L_{\rm inj}$ & $L_0$ & $N_0$ & RMS error & Max.\ abs.\ error \\
\hline
15 & 1 & 0 & 1  & $1.0\times 10^{-15}$ & $1.0\times 10^{-15}$ \\
15 & 1 & 1 & 4  & $1.8\times 10^{-3}$  & $4.9\times 10^{-3}$  \\
\hline
32 & 3 & 0 & 1  & $2.6\times 10^{-15}$ & $2.6\times 10^{-15}$ \\
32 & 3 & 1 & 4  & $6.0\times 10^{-7}$  & $3.9\times 10^{-6}$  \\
32 & 3 & 2 & 9  & $2.6\times 10^{-4}$  & $3.0\times 10^{-3}$  \\
32 & 3 & 3 & 16 & $6.7\times 10^{-3}$  & $9.8\times 10^{-2}$  \\
\hline
51 & 4 & 0 & 1  & $7.8\times 10^{-16}$ & $7.8\times 10^{-16}$ \\
51 & 4 & 1 & 4  & $5.5\times 10^{-5}$  & $1.9\times 10^{-4}$  \\
51 & 4 & 2 & 9  & $1.7\times 10^{-4}$  & $1.5\times 10^{-3}$  \\
51 & 4 & 3 & 16 & $9.2\times 10^{-4}$  & $2.3\times 10^{-2}$  \\
51 & 4 & 4 & 25 & $4.3\times 10^{-3}$  & $5.9\times 10^{-2}$  \\
\end{tabular}
\end{ruledtabular}
\end{table*}

Several features are evident in Table~\ref{tab:idaYlm}. First, the $L_0=0$ monopole sector is
reproduced to essentially machine precision for all three basis sizes. This is an important
baseline check, since the monopole channel is tied directly to the exact representation of the
constant function. Second, for $N=32$ and $N=51$ the errors remain extremely small in the
$L_0=1$ sector and are still modest through $L_0=2$, showing that the two-index IDA captures
the low-angular-momentum Coulomb structure very accurately. As $L_0$ approaches the top of the
injected block, the errors increase, which is expected: although the corresponding subspace is
represented exactly, the diagonal approximation itself is not. Finally, the denser $N=51$
basis improves significantly over $N=32$ at the higher tested values of $L_0$, while the
coarse $N=15$ basis is already accurate at the few-$10^{-3}$ level in the $L_0=1$ sector.

\bibliographystyle{apsrev4-2}
\bibliography{ref}

\end{document}